\documentclass[%
 reprint,superscriptaddress,amsmath,amssymb,aps
 ]{revtex4-1}
\usepackage{graphicx}
\usepackage{appendix}
\usepackage{dcolumn}
\usepackage{bm}
\usepackage{braket}
\usepackage{amsmath}
\usepackage{amssymb}
\usepackage{graphicx}
\usepackage[colorlinks=true, allcolors=blue]{hyperref}
\usepackage{mathtools}
\usepackage{enumerate}
\usepackage{textgreek}
\usepackage{caption}
\usepackage{subcaption}
\usepackage[font=small,
   justification=justified,
   format=plain]{caption}

\begin{document}
\title{Coherent amplification and lasing of weak surface-plasmon polaritons in a negative-index metamaterial with a resonant atomic medium}
\author{Saeid Asgarnezhad-Zorgabad}
\affiliation{Departement of physics, sharif university of technology, 11165-9161, Tehran, Iran}
\email{saeid.asgarnezhad@sharif.ir}
\date{\today}
\begin{abstract}
Surface plasmon polaritons (SPPs) lasing requires population inversion, it is inefficient and possesses poor spectral properties. We develop an inversion-less concept for a quantum plasmonic waveguide that exploits unidirectional superradiant SPP emission of radiation to produce intense coherent surface plasmon beams. Our scheme includes a resonantly driven cold atomic medium in a lossless dielectric situated above an ultra-low loss negative index metamaterial~(NIMM) layer. We propose generating unidirectional superradiant radiation of the plasmonic field within an atomic medium and a NIMM layer interface and achieve amplified SPPs by introducing phase-match between the superradiant SPP wave and coupled laser fields. We also establish a parametric resonance between the weak modulated plasmonic field and the collective oscillations of the atomic ensemble, thereby suppressing decoherence of the stably amplified directional polaritonic mode. Our method incorporates the quantum gain of the atomic medium to obtain sufficient conditions for coherent amplification of superradiant SPP waves, and we explore this method to quantum dynamics of the atomic medium being coupled with the weak polaritonic waves. Our waveguide configuration acts as a surface plasmon laser and quantum plasmonic transistor and opens prospects for designing controllable nano-scale lasers for quantum and nano-photonic applications.
\end{abstract}
\maketitle
\section{Introduction}
Surface-plasmon polariton lasers and amplifiers~\cite{berini2012surface}, also known as microscopic/nanoscopic sources of light are important for providing and modulating linear and nonlinear interactions within subwavelength scales~\cite{Holtfrerich:16,bogdanov2019overcoming}. These nanophotonic elements are valuable in designing quantum- and nonlinear-photonic technologies such as an SPP frequency-comb generator~\cite{PhysRevA.99.051802} phase rotors~\cite{Asgarnezhad_Zorgabad_2020} and quantum information processors~\cite{boltasseva2011low,tame2013quantum}. Recent material technologies for fabricating nanoplasmonic configurations~\cite{hess2012active} provide opportunity to exploit SPP lasing and amplifying~\cite{PhysRevLett.90.027402,noginov2009demonstration} in a wide range of applications such as in biology~\cite{galanzha2017spaser} and quantum generator~\cite{stockman2010spaser}. However, producing coherent lasing and stable amplification of SPPs is challenging due to the need for a giant phase mismatch for generating unidirectional SPP launching~\cite{zhang2019surface}, providing a population inversion in nano-steps~\cite{premaratne2017theory} and overcoming high Ohmic loss~\cite{oulton2009plasmon}.

On the other hand, lasing SPPs are inefficient~\cite{PhysRevLett.118.237402} and this amplification for plasmonic waves depends on the high laser powers~\cite{meng2013wavelength}, the dense concentration of the gain medium~\cite{noginov2009demonstration}, and well-designed nano-scale materials~\cite{lu2012plasmonic}. Introducing a high-input field commensurate with a low concentration rate of the active gain limits the amplification efficiency. Moreover, including a dense dipolar gain produces amplified spontaneous emission that reduces the surface plasmon lasing operation~\cite{meng2013wavelength}. Consequently, gain media properties and high-input field power induce noises to plasmonic systems that limit the efficiency of SPP lasers in the quantum regime. These limitations are challenging and prevent the realization of SPP lasing in an experiment~\cite{PhysRevLett.118.237402}.

Previous investigations demonstrate that the quantitative and qualitative descriptions of the plasmonic nanolaser~\cite{lu2012plasmonic} are possible for only quasi-static effects such as synchronizing plasmon oscillations with external injected field~\cite{Andrianov:11} and intensity-dependent frequency shifts~\cite{PhysRevA.86.043824}. These proposals indicate that the surface plasmon lasing is obtained for a nanoscopic dipole resonance or relaxations of the gain media~\cite{PhysRevLett.118.237402}. On the other hand, an investigation also reveals that quantum coherence can significantly enhance the surface plasmon amplification for a silver nano-particle that is coupled to the externally driven three-level gain medium~\cite{PhysRevLett.111.043601}. The presented experimental and theoretical schemes for realizing surface plasmon lasers are based on the stimulated emission of radiation, which is hard to achieve within nanoscopic scales.    

Ameliorating these limitations and developing a lasing scheme with low-intensity laser fields and without the need to population inversion, thereby provides the opportunity to exploit these nanoscopic sources of light as a coherent amplifier, fast modulators, and efficient nanolasers. By proposing an atomic ensemble and exciting a superradiant emission of radiation, it is shown that a weak probe field amplifies without the need for population inversion~\cite{PhysRevX.3.041001}. Recently, a proposal indicates that directional superradiant surface-plasmon polaritons can be launched in the interface between a graphene layer and a heralded atomic scheme~\cite{zhang2019surface}; however, the weak plasmonic field amplification is not investigated within the presented graphene plasmonic scheme.

Consequently, fundamental questions that may appear are whether SPPs can also be amplified without population inversion, whether amplification needs a high-power field, whether this intense SPP field is coherent and uni-directional, and what would be the spectral properties of this coherent amplification? We give affirmative answers to these open questions by devising an ultra-low loss quantum plasmonic scheme, that exploits SSPP emission for amplification of the weak SPP field. Our scheme introduces a quantum gain to a loss-compensated nanoscopic devise, which is novel and we explore its application to field-effect plasmonic transistors~\cite{sun2018single} and nanoscale quantum generators~\cite{PhysRevB.98.075411}.

\begin{figure}
    \includegraphics[width=\columnwidth]{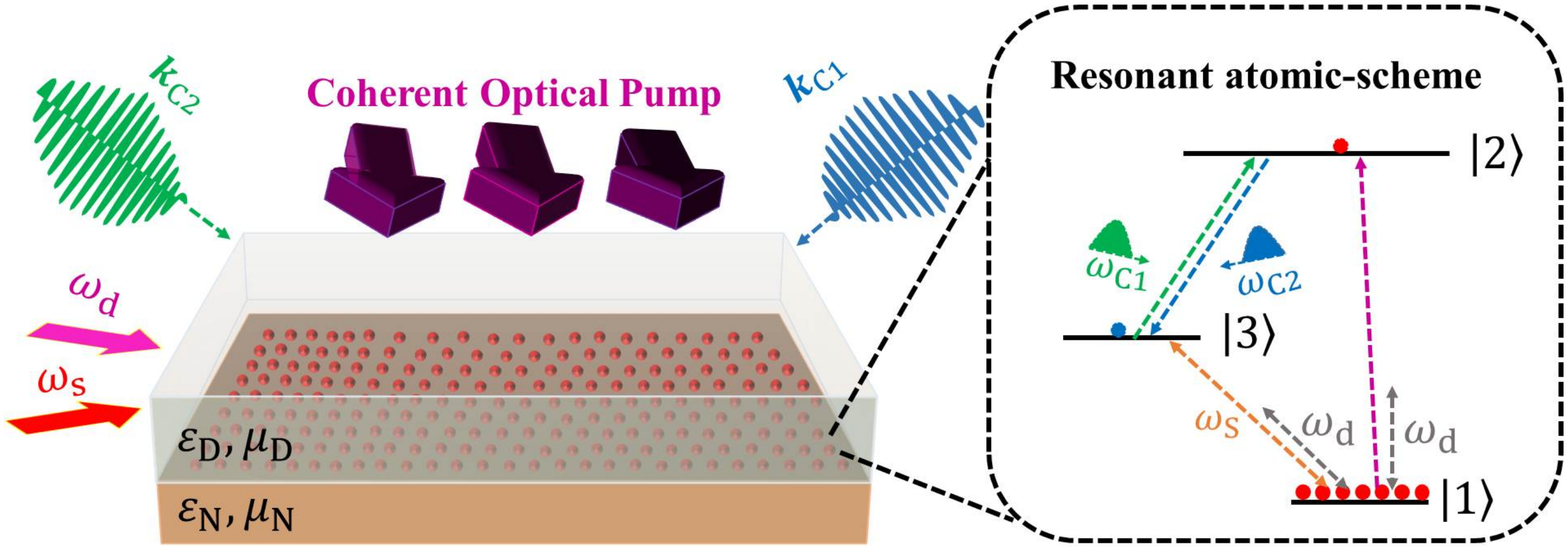}
\caption{\label{fig:Figm}%
Waveguide configuration
for the amplification of SSPP emission. The waveguide comprises a dielectric layer with electric permeability~$\varepsilon_\text{D}$ and a (NIMM) layer with optical constants~$(\varepsilon_\text{N},\mu_\text{N})$. The atomic medium is doped in the dielectric-NIMM layer interface. A coherent optical pump and a train couple laser fields provide phase matching, that yield SSPP excitation.}
\end{figure}
The rest of the paper is organized as follows: In \S~\ref{background} we present the background of our work. In \S~\ref{approach} we elucidate the quantitative description and experimental realization of our scheme, develop the mathematical analysis of our plasmonic amplifier, and present our methods to solve mathematical formalism. We explain the steps towards directional SSPP in \S~\ref{Sec:Superradiant_preperation} and we discuss the amplification of the weak SPP waves in the presence of the directional superradiant field in \S~\ref{Sec:Surface_Plasmon_Amplification}. Finally, we discuss and summarize our results in sections \S~\ref{Sec:Discussion} and \S~\ref{Sec:Conclusion}, respectively.   

\section{\label{background}Background}
We begin this section by briefly reviewing the superradiant emission of radiation in \S~\ref{Sec:Supereradiant_Background}. Next, we introduce the parametric resonance and discuss the possibility of providing gain for a weak driving field \S~\ref{Sec:PArameteric_Resonance}. Finally, we review the salient aspect of the weak field amplification using the Mathieu equation. This concept is discussed in \S~\ref{Sec:Weak_Field_Amplification_BAckground}.

\subsection{\label{Sec:Supereradiant_Background}Superradiant emission of radiation}
In this subsection, we elucidate the pertinent concept of the superradiant emission of radiation~\cite{PhysRev.93.99,PhysRevA.98.063815,RevModPhys.91.035003}. For a $N$ two-level atoms with ground state $\ket{b}$ and excited state $\ket{a}$ that are situated within a cell of radius $R$ much smaller than the radiation wavelength $\lambda$, a uniform absorption of photon by a single quantum emitter prepares the atomic ensemble to the so called Dicke-state
\begin{equation}
    \ket{\Psi_\text{s}}=\frac{1}{\sqrt{N}}\sum_{j=1}^{N}\ket{b_1,b_2,\ldots,a_j,\ldots,b_N}.
\end{equation}
This collectively excited atomic state decays into $\ket{b_1,b_2,\ldots,b_N}$ at a rate $\Gamma_\text{s}=N\gamma$; $\gamma$ is the single-atom decay rate, and consequently produce a superradiant emission~\cite{PhysRevX.3.041001}. On the other hand, wavevector of the propagated photon~($k_\text{a}$) would record through time-Dicke state
\begin{equation}
    \ket{\psi_\text{a}}=\frac{1}{\sqrt{N}}\sum_{j=1}^{N}\exp\{\text{i}\bm{k}_\text{a}\cdot\bm{r}_{j}\}\ket{b_1,b_2,\ldots,a_j,\ldots,b_N},
\end{equation}
if this single photon is uniformly absorbed by an atom situated on $\bm{r}_{j}$ within the atomic ensemble~\cite{PhysRevLett.96.010501}. Consequently, an atomic medium prepared to a time-Dicke state and situated above a metallic like layer, may produce a surface polaritonic superradiant emission through spontaneous decay to ground state and thereby produces a photon with wavevector $\hbar\bm{k}_\text{a}$ and energy $E=\hbar\omega_\text{ab}$~\cite{zhang2019surface}. In our work, directional SSPP launches for the wavenumber $\bm{k}_\text{SPP}=\bm{k}_\text{a}$ and perturbation frequency $\omega_\text{SPP}=\omega_\text{ab}$.

\subsection{\label{Sec:PArameteric_Resonance}Parametric resonance}
We start this subsection by introducing the concept of parametric resonance~\cite{PhysRevLett.91.111601,parametricResonance}. Parametric resonance is known as a process in which the parameters that describe a system possesses time variation or temporal evolution. In a physical configuration that is characterized with a periodic system parameter $a(t)$, and its dynamical evolution can be described with
\begin{equation}
    \ddot{X}+a(t)X=0,
    \label{Eq:General_Form_Matheiu}
\end{equation}
the parametric resonance occurs if
\begin{equation}
    a(t)=a(t+mT),
\end{equation}
for any positive integer $m\in\{2,3,4,\ldots\}$. It follows from Floquet’s theorem~\cite{PhysRevA.15.1109,CHU20041} that \eqref{Eq:General_Form_Matheiu} with a periodicity factor $\sigma$ has an arbitrary solution $X(t)$ such that
\begin{equation}
    X(t+2\pi)=\sigma X(t).
\end{equation}
The periodicity factor depends on the system parameters.

In this work, we employ the concept of parametric resonance to establish weak plasmonic field amplification in the presence of SSPP radiation. This amplification is achieved for a characteristic frequency $\omega_\text{ch}$ satisfying 
\begin{equation}
    \omega_\text{ch}=\frac{\omega_0}{m},
    \label{Eq:Parametric_Resonances_Background}
\end{equation}
for $\omega_0$ the frequency perturbation of the plasmonic field in which the amplification occurs.

\subsection{\label{Sec:Weak_Field_Amplification_BAckground}Weak field amplification and dynamical stability}
This subsection describes the key concepts of the weak field amplification and discusses the stability of the amplified field. Amplitude enhancement and amplification can be achieved within a dynamical system by exploiting \textit{forced} oscillations~\cite{kolmanovskii2012applied} and \textit{parametric} oscillations~\cite{10.1115/1.3641707}. In a physical system with two control parameter $a$, $q$ whose temporal dynamics is described with a specific form of \eqref{Eq:General_Form_Matheiu} known as Mathieu equation~
\begin{equation}
    \frac{\text{d}^2X}{\text{d}t^2}+(a-2q\cos(2t))X=0,
    \label{Eq:Mathieu_General_Background}
\end{equation}
parametric resonances \eqref{Eq:Parametric_Resonances_Background} would yield the field amplification. We notice that the strongest amplification is achieved for the first order of resonance~($m=1$)~\cite{PhysRevX.3.041001}. 

We achieve the solution of Eq.~\eqref{Eq:Mathieu_General_Background} in terms of arbitrary constants $A$, $B$, periodicity exponent $\mu$ and a periodic function $\Phi(t)$ by exploiting the Floquet's theorem
\begin{equation}
    X(t)=A\exp\{\mu t\}\Phi(t)+B\exp\{-\mu t\}\Phi(-t),
    \label{Eq:Solution_Matheiu_Background}
\end{equation}
that establishes stable amplification only for the specific values of $a$ and $q$~\cite{doi:10.1119/1.18290}. In this work, we establish that the dynamical evolution of the stable weak plasmonic field in the presence of SSPP describes by Mathieu-like equation~\eqref{Eq:Mathieu_General_Background} that can be amplified through parametric resonances.

\section{\label{approach}Approach}
In this section, we describe our plasmonic configuration in detail. First, we qualitatively model our waveguide in \S~\ref{Sec:Model} by introducing the source-waveguide-detection triplet. Next, in \S~\ref{Sec:Mathematic} we develop a quantitative approach to describe our system. We note that in elucidating the waveguide, we have employed Schr\"odinger equation to achieve the dynamics of the atomic medium, Drude-Lorentz model to describe the nano-fishnet metamaterial layer and we use Maxwell-Schr\"dinger equations to obtain the evolution of the weak SPP field. Finally, in \S~\ref{Sec:Method} we explain our quantitative approach to solve the resultant equations.

\subsection{\label{Sec:Model}Model}
In this subsection we qualitatively describe the plasmonic configuration, which comprises three parts, (i) source, (ii) waveguide and (iii) detection. Consequently, we begin our description by elucidate laser fields as source in \S~\ref{Sec:Source}, then we explain the waveguide configuration by describing the metamaterial layer and atomic medium in \S~\ref{Sec:Waveguide} and finally we describe the detection system for measuring the output spectral intensity of weak plasmonic field in \S~\ref{Sec:Detection}. Furthermore, we briefly discuss the possible realistic model of source-waveguide-detection triplet~\ref{Sec:Feasibility}.  

\subsubsection{\label{Sec:Source}Source}
The generation and amplification of the coherent SSPP field are obtained by exploiting four fields: a weak signal~(s), a strong driving~(d), a couple~(c) fields and an incoherent flash lamp. We assume our fields all share the same polarisation and are obtained from a dye laser that is frequency stabilized, linearly polarised, possesses enough spatial coherence to cover the waveguide and temporally longer than amplification scale~\cite{wang2008all}. Acousto-optic modulators control the carrier frequency of each beam. The fourth driving field is an optical pump from a flash lamp with long temporal width and linearly polarised, and spatially broadened enough to cover all the interaction surface~\cite{zhang2019surface}.

\subsubsection{\label{Sec:Waveguide}Waveguide}
Now we elucidate our plasmonic configuration. Laser fields as sources are injected on our planar waveguide, which comprises three layers: a substrate that a metamaterial can build on it, a double-negative-index metamaterial as middle layer and a dielectric on the top (see Fig.~\ref{fig:Figm}). On one ends of the waveguide an optical fiber is attached~\cite{Stegeman:83} and, on the other end, a Bragg grating structure~\cite{bonefacino2016recent}. The top of the NIMM layer, which could be constructed as a nano-fishnet structure~\cite{shalaev2007optical}, is doped by atoms or molecules serving as electric dipoles, and the depth of this dopant layer is a few dipole resonant wavelengths. The flash-light irradiates through the dielectric into the waveguide normal to the interface, signal and driving lasers are injected and co-propagate parallel to the interface with the end-fire coupling technique~\cite{PhysRevB.63.125417}. The couple field is separated into the two contra-propagate fields and introduced to the waveguide at a small angle normal to the interface. This source-waveguide-detector triplet is experimentally feasible and efficient for quantum SPP excitation~\cite{akimov2007generation}.

\subsubsection{\label{Sec:Detection}Detection}
Finally, we explain the operation of laser fields and explain our proposed detection scheme. The optical pump is applied to induce collective excitation of the atomic medium~\cite{bohnet2012steady,sonnefraud2010experimental}, the couple field provides an opportunity to generate directional SSPP~\cite{PhysRevLett.95.133601}, the strong driving field introduces a quantum gain to the hybrid plasmonic structure~\cite{PhysRevX.3.041001}, and the signal field produces a weak SPP field that we are supposed to amplify. These polaritonic waves propagate into the Bragg regime. This Bragg structure is a dielectric with modified optical properties. We employ a tapered multimode optical fiber due to its efficiency for detecting the intensity of the amplified SSPP. This fiber is suspended above the Bragg regime, which presumably is evanescently coupled to the Bragg regime of the waveguide. Certain spectral components are preferentially scattered. Spectral properties of the field propagating through the fiber are measured and used to infer spectral properties of the surface plasmon polariton.

\subsubsection{\label{Sec:Feasibility}Possible realistic model}
Specifically, we assume $\text{Pr}^{3+}$ ions doped in $\text{Y}_{2}\text{Si}\text{O}_{5}$ crystal with corresponding atomic levels
\begin{align}
 \ket1=&\ket{^2\text{H}_4,F=\pm\frac52},\; \ket2=\ket{^3\text{P}_0},\nonumber\\
 \ket3=&\ket{^2\text{D}_1,F=\pm\frac12}.
\end{align}
This medium has atomic density $N_\text{a}$, the natural decay rate for $\ket{m}\leftrightarrow\ket{n}$ is $\Gamma_{mn}$, dephasing rate is $\gamma_{mn}^{\text{deph}}$ and the scheme is also cooled up to $4~\text{K}$ and inhomogeneously broadened by $W_{ij}$~\cite{Ham:99} that affects the SSPP amplification. However, we consider spectral hole burning technique~\cite{moerner1988persistent}, thereby minimize the effect of these broadening on generated SPP and dynamics of weak SSPP. 

There are different metamaterial layers operate in the optical frequency region~\cite{shalaev2007optical,lazarides2018superconducting}. Our suggested NIMM layer is a nano-fishnet metamaterial. Specifically, we consider $\text{Al}_2\text{O}_3$-$\text{Ag}$-$\text{Al}_2\text{O}_3$ multi-layer with rectangular nano-holes as our NIMM, which possesses low Ohmic loss for the SPPs within optical frequencies~\cite{Xiao:09,Liang:17}.

Various mechanism such as optical parametric amplification~\cite{Popov:06}, geometrical tailoring and optimization~\cite{PhysRevB.80.125129}, including gain media~\cite{xiao2010loss} and meta-surfaces~\cite{Genevet:17} for which the Ohmic loss related to plasmonic structures, specifically for optical frequencies can be effectively reduced. However, we introduce a \textit{virtual} gain to our hybrid interface by employing a coherent method, which is based on constructive interference of the two externally injected plasmonic fields and also excited signal SPP field to suppress the Ohmic loss of this NIMM layer~\cite{PhysRevLett.115.035502,Ghoshroy:20}. This loss-compensation is basically different from the stimulated optical loss suppression achieved by dye molecules or amplification induced in integrated plasmonic chip~\cite{xiao2010loss,fedyanin2012surface}.

\subsection{\label{Sec:Mathematic}Mathematical description of the polaritonic waveguide}
This subsection starts with the quantitative description of the waveguide. First, we describe NIMM and develop the mathematical formalism to describe our metamaterial layer. Next, we describe the atomic medium and elucidate its interaction with metamaterial layer by introducing the interactive Hamiltonian commensurate with the Schr\"odinger approach and we finally present a mathematical description to elucidate the confinement of surface polaritonic field to the interaction interface.

To characterize the optical properties of the NIMM layer, we employ macroscopic description of the permittivity~($\varepsilon_\text{N}$) and permeability~($\mu_\text{N}$) following Drude-Lorentz model~\cite{PhysRevLett.101.263601,Sang_Nourpour_2017}. To this aim, we introduce the permittivity
\begin{equation}
    \varepsilon_\text{N}=\varepsilon_{\infty}-\frac{\omega_\text{e}^2}{\omega(\omega+\text{i}\gamma_\text{e})}
    \label{Eq:Permittivity_NIMM}
\end{equation}
and permeability
\begin{equation}
    \mu_\text{N}=\mu_{\infty}-\frac{\omega_\text{m}^2}{\omega(\omega+\text{i}\gamma_\text{m})},
    \label{Eq:Permeability}
\end{equation}
for $\varepsilon_{\infty}$
and $\mu_{\infty}$ the background constant for the permittivity and permeability, respectively, $\omega$ the perturbation frequency, $\omega_\text{e}$ and~$\omega_\text{m}$ are the electric and magnetic plasma frequencies, and $\gamma_\text{e}$ and~$\gamma_\text{m}$ are the corresponding decay rates~\cite{Asgarnezhad_Zorgabad_2020}.

Next, we evaluate the interaction Hamiltonian of this plasmonic scheme in three steps: (i) employ canonical quantization method, by introducing bosonic creation/annihilation operators to achieve the quantized current density within quantum emitter-NIMM layer interface, (ii) use this current to quantized the electric field component of SPP mode, and (iii) exploit a dipole approximation and express the Hamiltonian of the system in terms of our bosonic operators and atomic dipole moment.

We evaluate quantized current density $\hat{\bm{j}}(\bm{r},\omega)$ in the interface between a dielectric and our NIMM layer interface by considering
\begin{equation}
    \mu_\text{N}(\bm{r},\omega):=[\kappa_\text{N}(\bm{r},\omega)]^{-1},
\end{equation}
and defining 
\begin{align}
    \alpha(\bm{r},\omega)=&\left\{\frac{\hbar\varepsilon_0}{\pi}\text{Im}[\varepsilon_\text{N}(\bm{r},\omega)]\right\}^{1/2},\label{Eq:Definition_Permittivity}\\
    \beta(\bm{r},\omega)=&\left\{-\frac{\hbar}{\pi\mu_{0}}\text{Im}[\kappa_\text{N}(\bm{r},\omega)]\right\}^{1/2},
    \label{Eq:Definition_Permeability}
\end{align}
as~\cite{Philbin_2010,Horsley_2014}
\begin{align}
    \hat{\bm{j}}(\bm{r},\omega)=&-2\pi\text{i}\omega\alpha(\bm{r},\omega)\hat{\bm{C}}_\text{e}(\bm{r},\omega)\nonumber\\&+2\pi\bm{\nabla}\times\left\{\beta(\bm{r},\omega)\hat{\bm{C}}_\text{m}(\bm{r},\omega)\right\}.
    \label{Eq:Quantized_Current_Density}
\end{align}
Here we assume $\hat{\bm{C}}_{\jmath}(\bm{r},\omega)$~($\hat{\bm{C}}_{\jmath}^{\dagger}(\bm{r},\omega))$; $\jmath\in\{\text{e},\text{m}\}$ as annihilation~(creation) operators associated with the electrical~(e) and magnetic~(m) response of the medium, whose components are described by usual bosonic commutation relation
\begin{align}
    \left[\hat{C}_{\jmath i}(\bm{r},\omega),\hat{C}_{\jmath' j}(\bm{r}',\omega')\right]=&0,\label{Eq:Commutation_Relation_First}\\
    \left[\hat{C}_{\jmath i}(\bm{r},\omega),\hat{C}_{\jmath' j}^{\dagger}(\bm{r}',\omega')\right]=&\delta_{ij}\delta_{\jmath\jmath'}\delta(\omega-\omega')\delta(\bm{r}-\bm{r}'). \label{Eq:Commutation_Relation_Second}
\end{align}
We calculate the quantized current density for our specific NIMM layer by plugging Eqs.~\eqref{Eq:Permittivity_NIMM} and \eqref{Eq:Permeability} into Eqs.~\eqref{Eq:Definition_Permittivity} and \eqref{Eq:Definition_Permeability}.

Next, we employ quantized current density characterized by Eq.~\eqref{Eq:Quantized_Current_Density} to evaluate the quantized electric field operator within interaction interface. This electric field is related to the Dyadic green function at the interface $\mathcal{A}(\bm{r},\bm{r}';\omega)$. In the limiting case of dielectric-metamaterial interface, this green function can be calculated similar to Ref.~\cite{PhysRevA.84.053824}. To this aim, first we define 
\begin{equation}
    \int_{\bm{r},\tilde{\omega}}:=\text{i}\frac{\mu_{0}}{2\pi}\int\text{d}^{3}\bm{r}\int_{0}^{\infty}\text{d}\tilde{\omega},
\end{equation}
and then express the quantized electric field in terms of system parameters as
\begin{equation}
    \bm{E}(\bm{r},t)=\int_{\bm{r}',\tilde{\omega}}\hbar\tilde{\omega}\left[\bm{\mathcal{A}}(\bm{r},\bm{r}';\omega)\cdot\hat{\bm{j}}(\bm{r}',\tilde{\omega})\text{e}^{\text{i}\tilde{\omega}t}+\text{h}.\text{c}\right].
    \label{Eq:Quantized_Electric_Field_Text}
\end{equation}
We employ this quantized electric field to describe the interaction Hamiltonian for the interface between the NIMM layer and the atomic medium interface. 

Finally we employ the quantized electric field, commensurate with bosonic annihilation/creation operators to evaluate the total Hamiltonian of the system. To this aim, we define the atomic energy levels by
\begin{equation}
    E_j=\hbar\omega_{j},\;\;\;\; j\in\{1,2,3\}
\end{equation}
for each atomic state $\ket{j}$. We assume the dipole moment of the $\ket{3}\leftrightarrow\ket{1}$ as $\bm{d}_{l}$ and also introduce the Pauli matrices correspond to this atomic medium as
\begin{align}
    \sigma_{l}^{z}=&\ket{3_{l}}\bra{3_{l}}-\ket{1_{l}}\bra{1_{l}},\\
    \sigma_{l}^{x}=&\ket{3_{l}}\bra{1_{l}}+\ket{1_{l}}\bra{3_{l}}.
\end{align}
The Hamiltonian of this plasmonic scheme then becomes
\begin{align}
    H=&\sum_{l=1}^{N_\text{a}}\left[\frac{\hbar\omega}{2}\sigma_{l}^{z}-\sigma_{l}^{x}\bm{d}_{l}\cdot\bm{E}_{l}(\bm{r}_{l})\right]\nonumber\\&+\sum_{\jmath}\int\text{d}^3\bm{r}\int_{0}^{\infty}\text{d}\tilde{\omega}\hbar\tilde{\omega}\hat{\bm{C}}_{\jmath}^{\dagger}(\bm{r}',\omega)\cdot\hat{\bm{C}}_{\jmath}(\bm{r}',\omega).
    \label{Eq:Total_Hamiltonian}
\end{align}
To achieve the dynamics of the atomic medium we employ Schr\"odinger equation
    \begin{equation}
    \frac{\partial\ket{\Psi(t)}}{\partial t}=-\frac{\text{i}}{\hbar}H\ket{\Psi(t)},
    \label{Eq:Schrodinger_Equation}
\end{equation}
introduce the time-dependent amplitudes $c_{3(1)}(t)$ to $\ket{3}$~($\ket{1}$) transitions and assume the Ansatz $\ket{\Psi(t)}$ as
\begin{align}
    \ket{\Psi(t)}=&\sum_{l=1}^{N_\text{a}}c_{3,l}\ket{3_{l},\phi}\otimes_{j\neq l}\ket{1_{j}}\nonumber\\&+\sum_{\jmath,m}\int\text{d}^3\bm{r}\int_{0}^{\infty}\text{d}\tilde{\omega}c_{1j}(\tilde{\omega},\bm{r}')\ket{g,\bm{I}_{\jmath, m}},
    \label{Eq:Initial_Ansatz}
\end{align}
where we assume $\ket{\phi}=\ket{g_1,g_2,\ldots,g_N}$ as a ground state of the atomic medium and
\begin{equation}
    \ket{\bm{I}_{\jmath, m}}=\hat{C}_{\jmath m}^{\dagger}(\bm{r}',\omega)\ket{\phi},
\end{equation}
 as the excited plasmonic mode within the interface, respectively. We achieve the dynamical evolution of the SSPP by solving Eq.~\eqref{Eq:Schrodinger_Equation} commensurate with \eqref{Eq:Initial_Ansatz} when the atomic medium is prepared to a time-Dicke state.

\subsection{\label{Sec:Method}Method}
In this work, dynamics of SSPP in the interface between the atomic medium and nano-fishnet metamaterial layer and stable amplification of the weak plasmonic wave without need to population inversion are obtained by employing the multiple scaled time and asymptotic expansion to Mathieu equation and Fourier optics of surface polaritonic wave. Consequently, first, we describe our perturbation technique by elucidating asymptotic expansion commensurate with multiple scale variables in \S~\ref{Sec:Multiple_Scale} and next, we briefly discuss the Fourier optics of SPP waves in \S~\ref{Sec:Fourier_Optics}. Finally, we use these mathematical techniques and employ the concepts presented in background to describe the amplification of the weak surface polaritonic field in the presence of directional of SSPP radiation.

\subsubsection{\label{Sec:Multiple_Scale}Multiple-scale variable and asymptotic expansion}
Our methods for solving the Mathieu/Hill equations within this hybrid plasmonic waveguide is based on multiple scale time variable commensurate with the asymptotic expansions~\cite{nayfeh2008perturbation}. To this aim, first we define the order of perturbation and then express this concept for multiple expansion of any arbitrary function.

To define the order of perturbation, let us assume $\varphi_{n}:\mathbb{R}\setminus 0\rightarrow\mathbb{R}$ as a sequence functions for $x\rightarrow 0$ if for any $n\in\{1,2,3,\ldots\}$ there is a gauge function $\varphi_n$ satisfying~\footnote{We write $f(\epsilon)=o[g(\epsilon)]$ as $\epsilon\rightarrow0$ if for any positive number $\delta$, independent of $\epsilon$, there exist $\epsilon_0$ such that $|f(\epsilon)|\leq\delta|g(\epsilon)|$ for $|\epsilon|\leq|\epsilon_0|$~\cite{nayfeh2008perturbation}.}
\begin{equation}
    \varphi_{n+1}=o(\varphi_{n}).
    \label{Eq:Gauge_Field}
\end{equation}
Then asymptotic expansion of any arbitrary function $f(x)$ and characterized $N\in\mathbb{N}$ in terms of these sequence functions is defined as a serious
\begin{equation}
    f(x)\sim\sum_{n=0}^{N}\epsilon^{n}\varphi_{n},
\end{equation}
only if 
\begin{equation}
    f(x)-\sum_{n=0}^{N}\epsilon^{n}\varphi_{n}=o(\varphi_{N}).
    \label{Eq:Condition_Asymptotic_First}
\end{equation}
We can rewrite \eqref{Eq:Condition_Asymptotic_First} as~\footnote{We write $f(\epsilon)=\mathcal{O}[g(\epsilon)]$ as $\epsilon\rightarrow0$ if for any positive number $A$, independent of $\epsilon$, there exist $\epsilon_0>0$ such that $|f(\epsilon)|\leq A|g(\epsilon)|$ for $|\epsilon|\leq|\epsilon_0|$~\cite{nayfeh2008perturbation}.}
\begin{equation}
    f(x)-\sum_{n=0}^{N-1}\epsilon^{n}\varphi_{n}=\mathcal{O}(\varphi_{N}).
    \label{Eq:Condition_Asymptotic_Second}
\end{equation}
Now, in this work we employ this concept to asymptotically expand the weak signal field Rabi frequency~($\Omega_\text{s}$) and solve the resultant Mathieu differential equation perturbatively to achieve the sufficient condition for plasmonic field amplification without the need to population inversion.

\subsubsection{\label{Sec:Fourier_Optics}Fourier optics of surface plasmons}
Our method for calculating quantized electric field is based on green function method~\cite{Philbin_2010,Horsley_2014} that is a Dyadic tensor. This Dyadic green function for surface polaritonic waves can be obtained in two different mechanism, namely (i) real frequency~($\omega$) and complex wavenumber $k$, and (ii) complex frequency and real wavenumber~\cite{PhysRevB.79.195414}. For a characterized plasmonic green tensor within a dissipative hybrid interface, the quantized electric field is then related to the green tensor with Eq.~\eqref{Eq:Quantized_Electric_Field_Text}. We assume relative position $\delta\bm{r}=\bm{r}-\bm{r}'$ and relative time $\delta t=t-t'$ to express this green function in a Fourier space as
\begin{equation}
    \bm{\mathcal{A}}(\bm{r},\bm{r}';\omega)=\int\frac{\text{d}^2\bm{q}}{(2\pi)^2}\int\frac{\text{d}\tilde{\omega}}{2\pi}\bm{g}(\bm{q},z,z';\tilde{\omega})\text{e}^{\text{i}[\bm{q}\cdot\delta\bm{r}-\omega\delta t]}.
    \label{Eq:Fourier_Spectrum_Green_Function}
\end{equation}
We interpret \eqref{Eq:Fourier_Spectrum_Green_Function} as the general Dyadic green tensor for a dissipative interface and we describe the propagation properties of the SSPP and dynamical evolution of the weak plasmonic field using \eqref{Eq:Fourier_Spectrum_Green_Function}.

Aforementioned explanation is valid for a propagating surface-plasmonic wave in our hybrid interface. Consequently, our quantum SPP should also be described using complex wavenumber or complex frequency representations in the Fourier space. This representation would depend on the zeros of the propagation constant. Now we present this explanation in mathematical details. To this aim, fist, we note that the SPP field dispersion in the interface between a dielectric and a NIMM layer is
\begin{equation}
    q(\omega_l)=\frac{\omega_l}{c}\sqrt{\frac{\varepsilon_\text{N}\varepsilon_0(\varepsilon_\text{N}\mu_0-\varepsilon_0\mu_\text{N})}{\varepsilon_\text{N}^2-\varepsilon_0^2}}.
\end{equation}
The roots correspond to this propagation constant~(i.e. $q(\omega_l)=0$) in our interactive interface is achieved in two alternative mechanisms, (i) considering $\omega_l$ as a real parameter to find a complex root for wavenumber, or (ii) introducing a real value $\bm{q}$ to find the complex root of perturbation frequency. Consequently, the green tensor related to this SPP dispersion can also be evaluated in terms of these considerations.

Next, we employ residues theorem~\cite{brown2009complex} to calculate the plasmonic green function for roots in complex $\tilde{\omega}$-real $|\bm{q}|$ Fourier space. We assume $\omega_\text{SPP}\approx\omega_{31}$ as the frequency excitation of our plasmonic mode. Now, the plasmonic green tensor can be represented for complex $\bm{q}$ or complex $\tilde{\omega}$. The plasmonic green tensor~($\bm{\mathcal{A}_\text{SPP}(\bm{q},z,z')}$) corresponds to this complex SPP frequency excitation $\omega_\text{SPP}$ and $-\omega_\text{SPP}^{*}$ becomes
\begin{equation}
    \bm{g}(\bm{q},z,z';\tilde{\omega})=\frac{\mathcal{A}_\text{SPP}(\bm{q},z,z')}{\tilde{\omega}-\omega_\text{SPP}}+\frac{\mathcal{A}_{-\text{SPP}}(\bm{q},z,z')}{\tilde{\omega}+\omega_\text{SPP}^{*}},
\end{equation}
for a real $q$ space, whose components are characterized by $q_x,q_y$. Alternatively, by considering real $\omega_{l}$ and complex wavenumber $\bm{q}$, this green tensor can be achieved in terms of characteristic complex SPP wavenumber $\bm{k}_\text{SPP}$ and $-\bm{k}_\text{SPP}^{*}$ as
\begin{equation}
    \bm{g}(\bm{q},z,z';\tilde{\omega})=\frac{\mathcal{A}_{q_{x}}(q_{z},z,z';\tilde{\omega})}{q_{x}-k_\text{SPP}}+\frac{\mathcal{A}_{-q_{x}^{*}}(q_{z},z,z';\tilde{\omega})}{q_{x}+k_\text{SPP}^{*}}.
\end{equation}
In this work, we develop our method and calculate the green tensor of the atomic medium-NIMM layer interface based on surface plasmon Fourier optics~\cite{PhysRevB.79.195414}.

\section{\label{Sec:Results}Results}
We present the main results of this paper in four sections: First, we give a qualitative description of directional SSPP propagation and directional SPP lasing operation in \S~\ref{Sec:Qualitative_description}. Second, in \S~\ref{Sec:Superradiant_preperation} we discuss the directional launching of SSPP. Next, in \S~\ref{Sec:Surface_Plasmon_Amplification} we present the details  weak SPP field amplification in the presence of this directional plasmonic superradiant radiation. Finally in \S~\ref{Sec:Detection_Amplified} we suggest a technique to detect directional amplified surface polaritonic wave.

\subsection{\label{Sec:Qualitative_description}Qualitative description of SSPP launching and SPP lasing operation}
Our hybrid plasmonic interface is inherently dissipative, and consequently propagation length and stability of the excited SSPP would be highly limited due to high loss. However, we suppress the Ohmic loss related to the NIMM layer by inducing the concept of virtual gain. This configuration then is suitable for both directional SSPP launching and stable propagation of SPP field, thus provides opportunity to coherent amplification of weak plasmonic field by generating polaritonic superradiant. Consequently, for coherent amplification of the weak SPP field, first, we discuss directional plasmonic superradiant excitation and then we propose amplification of the weak SPP field. In what following, we qualitatively elucidate the main steps towards launching SSPP excitation, then as a second step we explain the give a brief discussion on weak plasmonic field amplification. 

\emph{Launching directional SSPP}- Directional SSPP within interaction interface is achieved by collective atomic ensemble excitation through two steps. First we perpendicularly illuminate the interaction interface using an optical pump with intensity $I_\text{p}$ to excite a single atom through coupling with $\ket{1}\leftrightarrow\ket{2}$ transition. Second, the contra-propagating couple laser fields with $(2n_\text{p}+1)\pi$ pulses and alternating wavevectors $\bm{k}_{\text{C}\imath}$; $\imath\in\{1,2\}$ drive the $\ket{2}\leftrightarrow\ket{3}$ atomic transition. The spontaneous emission of $\ket{3}\leftrightarrow\ket{1}$ transition for the atom in position $\bm{r}_{j}$ consequently generates a single SPP mode with characteristic wavenumber~\cite{zhang2019surface}
\begin{equation}
  \bm{k}_\text{SPP}=(n_\text{p}+1)\bm{k}_{\text{C}2}-n_\text{p}\bm{k}_{\text{C}1},  
  \label{Eq:Phase_Match}
\end{equation}
and this directional plasmonic emission serves as directional superradiant mode due to the atomic medium being prepared as~\cite{PhysRevLett.96.010501,PhysRevLett.113.083601}
\begin{equation}
    \ket{\psi_\text{SPP}}
    =\frac{1}{\sqrt{N_\text{a}}}
    \sum_{i=1}^{N_\text{a}}\exp\left[\text{i}
    \bm{k}_\text{SPP}\cdot \bm{r}_i\right]
    \ket{3_{i}}\otimes_{j\neq i}\ket{1_{j}}.
    \label{timed-dicke state}
\end{equation}
Our proposed Dicke-state with wavenumber $k_\text{SPP}$ and frequency $\omega_\text{SPP}$ decays faster than the single atom spontaneous emission, thereby acts as directional superradiant emission~\cite{PhysRevLett.96.010501,PhysRevX.3.041001}.  

\emph{Weak plasmonic field amplification}- We describe the coherent amplification of the weak SPP wave in five steps. First, we employ Schr\"odinger equation formalism to introduce a directional SSPP emission between $\ket{3}\leftrightarrow\ket{1}$ transition~\cite{PhysRevLett.96.010501} within the atomic medium-NIMM layer interface. Second, we assume the weak signal SPP field strongly coupled to the interface with a evanescence function, propagate as a traveling wave and possesses a constant phase whose spatiotemporal dynamics is achieved by using the coupled Maxwell-Schr\"dinger commensurate with its Fourier spectrum. Next, we employ a coherent loss-compensation mechanism and quantum decoherence suppression by using the resonant coupling between the collective atomic excitation and plasmonic fields. Finally, we investigate the coherent amplification commensurate with stability analysis of this weak plasmonic wave by introducing the parametric resonance to this interface. Consequently, to quantitative description of the system, we employ three well-established assumptions, namely (i) Schr\"odinger equation to launch directional SSPP, (ii) Drude-Lorentz model to describe nano-fishnet NIMM layer, and (iii) Maxwell-Schr\"odinger equation to achieve spatiotemporal dynamics of weak signal SPP.

\emph{Vision for our quantitative description}- We express our results by presenting qualitative and quantitative descriptions towards weak plasmonic field amplification. Note that our quantitative description for obtaining weak SPP amplification is based on three main equations, namely, (i) dynamics of the excited atomic state, Eq.~\eqref{Eq:Volterra}, (ii) dynamics of the weak SPP field in the interaction interface, Eq.~\eqref{Eq:Reduced_Maxwell_Equation}, and (iii) weak-field amplification through Mathieu-like equation \eqref{Eq:Matuie_Equation}. First, we present main mathematical steps for calculating the Hamiltonian of the system and then employ Sch\"odinger approach to describe the dynamics of excited state and launching directional SSPP in \S~\ref{Sec:Mathematical_Formalism}, next, we employ Fourier optics of SPP wave to represent the main steps toward derivation of \eqref{Eq:Reduced_Maxwell_Equation} in \S~\ref{Sec:Mathematical_Formalism_Amplificaiton} and finally we provide main steps to evaluate Mathieu-like equation in \S~\ref{Sec:Dynamical_Evolution_Amplified_SPP}.

\subsection{\label{Sec:Superradiant_preperation}Superradiant surface-plasmon polariton launching}
We present the excitation of directional SSPP in our hybrid plasmonic waveguide in two steps: First, we elucidate the mathematical formalism towards SSPP dynamics in \S~\ref{Sec:Mathematical_Formalism} and next in \S~\ref{Sec:Dynamical_Evolution_SSPP} we explain the dynamical evolution of the atomic excited state and establish the launching the polaritonic superradiant radiation within our interaction interface.

\subsubsection{\label{Sec:Mathematical_Formalism}Mathematical formalism of superradiant launching}
In this section, we develop our mathematical formalism towards directional plasmonic superradiant radiation. Based on our qualitative description, we achieve SSPP by preparing the atomic medium to a time-Dicke state characterized by \eqref{timed-dicke state}. To efficient excitation of SSPP, we suggest the heralded atomic ensemble~\cite{scully2009super} due to its efficiency for generating superradiant pulse. We explain launching SSPP for a simple case where the driving~(d) and signal~(s) fields are switched off~($\Omega_\text{d}=\Omega_\text{s}=0$). In this case, coupling optical pump and $(2n_\text{p}+1)\pi$ couple pulses as we describe in \S~\ref{Sec:Qualitative_description} would yield excitation of SSPP.

We describe this quantum plasmonic excitation by exploiting a quantized electric field $\hat{\bm{E}}$ characterized by \eqref{Eq:Quantized_Electric_Field_Text}, a quantized current density $\hat{\bm{j}}$ that is obtained by \eqref{Eq:Quantized_Current_Density}~\cite{Philbin_2010}. Moreover, to obtain quantized electric field, we need the plasmonic tensor. Following the calculation represented in~\cite{PhysRevB.79.195414} and \S~\ref{Sec:Fourier_Optics} we also consider the green tensor of the system as
\begin{align}
      \mathcal{A}_\text{SPP}=&\text{i}\frac{k_\text{a}\varepsilon_\text{N}}{k_0\sqrt{\varepsilon_\text{a}}}\frac{C(k,\tilde{\omega})}{2\text{Re}[\omega_\text{SPP}]}\nonumber\\&\times\left[\bm{e}_{k}-\frac{k}{k_\text{N}}\bm{e}_{z}\right]\hat{\bm{u}}_\text{SPP}(\bm{r}_{l})\text{e}^{\text{i}(k_\text{N}-k_\text{a})z}, 
  \label{Eq:Green_Funtion_Interface}
\end{align}
for $\varepsilon_\text{a}$ the permittivity of atomic medium, $\hat{\bm{u}}_\text{SPP}$ the unit vector of the excited plasmonic field and with
\begin{equation}
    k_\text{a}^{2}=k_\text{SPP}^{2}-\tilde{\omega}^{2}\mu_\text{a}\varepsilon_\text{a}^{2}/c^{2}.
\end{equation}
Our calculated quantized electric field and current density within hybrid atomic medium-NIMM layer interface are then described by annihilation-creation operators \eqref{Eq:Commutation_Relation_First} and~\eqref{Eq:Commutation_Relation_Second} that obey bosonic commutation relation~\cite{horsley2014canonical,PhysRevB.82.075427,PhysRevA.52.4823}. By evaluating the quantized electric field commensurate with the commutation relation, we characterized the Hamiltonian of the system \eqref{Eq:Total_Hamiltonian}.

Now, we investigate the dynamical evolution of the atomic medium within our interaction interface. We substitute the Hamiltonian of the system \eqref{Eq:Total_Hamiltonian} and the atomic Ansatz as~\eqref{Eq:Initial_Ansatz} to the Schr\"odinger equation \eqref{Eq:Schrodinger_Equation}~\cite{scully1999quantum}. To solve the resultant equations, we employ mapping
\begin{equation}
    \sum_{\jmath m}\int\text{d}^{3}\bm{r}\int_{0}^{\infty}\text{d}\tilde{\omega}\mapsto\int_\text{SPP},
\end{equation}
and we define the emitter-emitter coupling~(for two characterized atom a, b) as
\begin{equation}
    \frac{\mu_0\omega^2}{\pi}\bm{d}_\text{a}\cdot\mathcal{A}\cdot\bm{d}_\text{b}\sim g_\text{ab}.
\end{equation}
We then achieve the temporal evolution of the atomic medium as
\begin{align}
    \frac{\partial c_{1}(t)}{\partial t}=&\text{i}\sum_{l=1}^{N_\text{a}}c_{3}(t)g_\text{ab}^{*}(\bm{r}_\text{a},\bm{r}';\tilde{\omega})\text{e}^{\text{i}(\tilde{\omega}-\omega_{31})t},\label{Eq:Lower_state}\\
    \frac{\partial c_{3}(t)}{\partial t}=&\text{i}\int_\text{SPP}c_{1}(\bm{r}',\tilde{\omega})g_\text{ab}(\bm{r}_\text{a},\bm{r}';\tilde{\omega})\text{e}^{-\text{i}(\tilde{\omega}-\omega_{31})t}.\label{Eq:Excited_state}
\end{align}
Next we perform a time integration of Eq.~\eqref{Eq:Lower_state} and substitute the resultant equation in Eq.~\eqref{Eq:Excited_state}, and introduce the dispersion and dissipation due to plasmonic field as
\begin{equation}
    \xi(\bm{q}):=\braket{\bm{q}|\bm{L}}:=\text{e}^{\text{i}\bm{q}\cdot\bm{L}}.
\end{equation}
Now, we investigate the dynamics of the $\ket{3}$ atomic state that is corresponds to the temporal evolution of directional SSPP.

We present the remaining steps of derivation in appendix~\ref{Section:Dynamics_Superradiant} and establish that the evolution of this excited state for a $j$th atom in a fixed position $\bm{r}_{j}$ is
\begin{equation}
    -\frac{\partial c_{3}(t)}{\partial t}
    =\omega_\text{o}^2\exp\{(t_\text{L}\gamma_\text{SPP})^2\}\int_0^{t}\text{d}t'\;K(t,t')c_{3}(t'),
    \label{Eq:Volterra}
\end{equation}
with
\begin{equation}
    t_L=L/v_\text{SPP},
\end{equation}
the SPP-flight time, $\tau_\text{l}=\gamma_\text{SPP}^{-1}$ is the time related to loss,
\begin{equation}
    \omega_\text{o}^2:=[N_\text{a}/(2L)^2]g_{ab},
\end{equation}
the oscillation frequency of the quantum SPP and with
\begin{equation}
    K(t,t')=\exp\left\{\frac{-(t-t'+2t_L^2\gamma_\text{SPP})^2}{(2t_L)^2}\right\},
    \label{Eq:Kernel_Volterra_Appendix}
\end{equation}
the kernel of~\eqref{Eq:Volterra}. Numerical solution of this equation then describes the dynamical evolution of the SSPP field. Note that Eq.~\eqref{Eq:Volterra} is an integro-differential equation that we solve it numerically. This equation is a boundary value problem with certain initial condition. In our configuration, we achieve plasmonic superradiant emission by preparing the atomic medium to a time-Dicke state. Consequently, in solving Eq.~\eqref{Eq:Volterra} we assume $c_{3}(t=0)=1$ and we neglect the temporal evolution of this excited state $\dot{c}_{3}(t=0)=0$.

\subsubsection{\label{Sec:Dynamical_Evolution_SSPP}Dynamical evolution of plasmonic superradiant radiation}
\begin{figure}
    \centering
    \includegraphics[width=1\columnwidth]{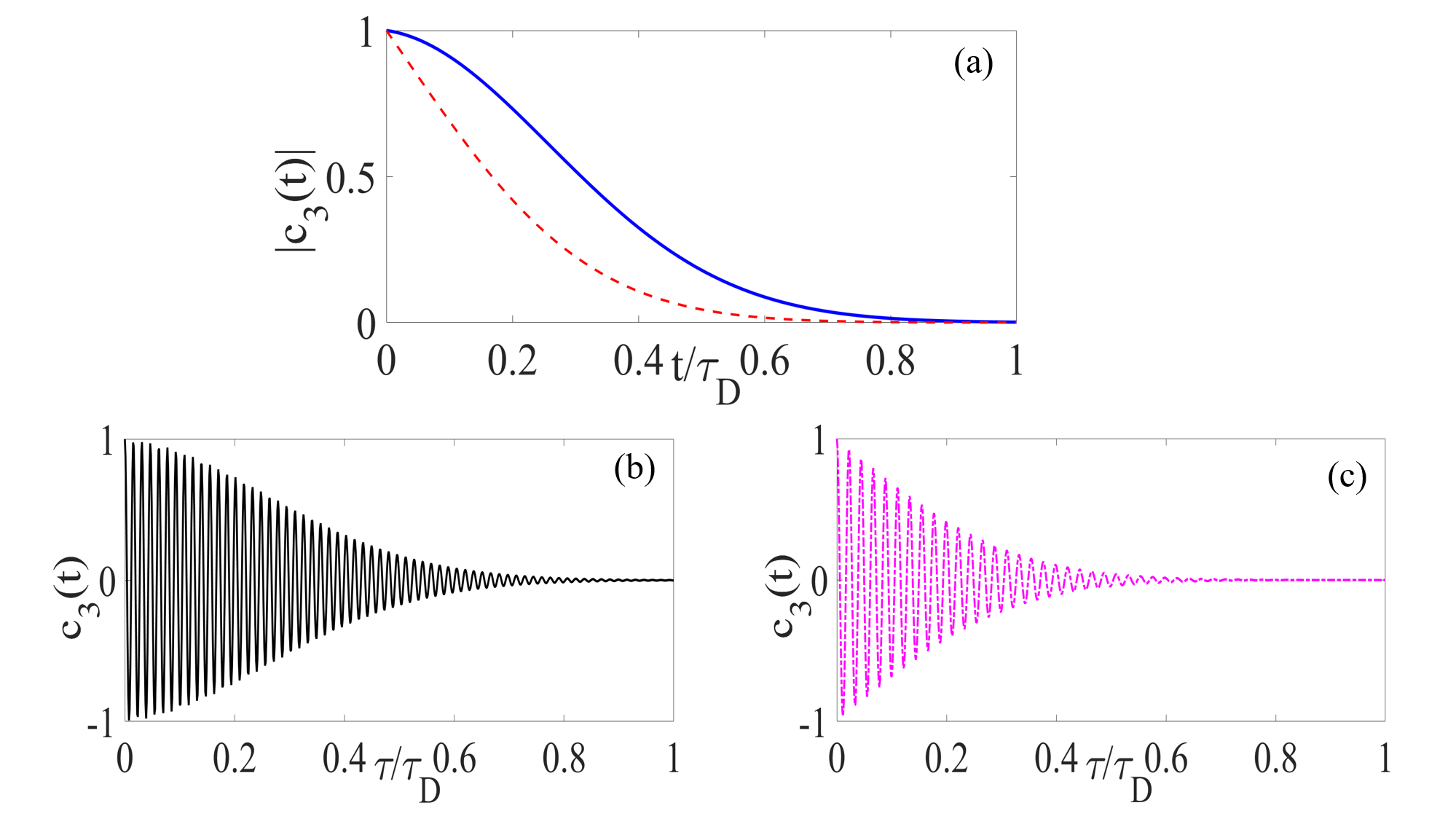}
    \captionof{figure}{Panel (a) denotes the bound solution of our integrodifferential equation. Panels (b) and (c) represent the numeric solution of the Eq.~\ref{Eq:Volterra}. In both panels, the solid lines (blue and black) represent the dynamics of the~$\ket3$ states for NIMM layer with coherent loss-suppression, while the dashed-line (red and magenta) curves are the results with the simple nano-fishnet metamaterial structure.}
    \label{fig:two}
\end{figure}

First, we investigate the dynamical evolution of the atomic states in Fig.~\ref{fig:two}. The evolution of the atomic ensemble through the time-Dicke state and the resultant collective oscillation would directly proportional to the SSPP emission in the Fourier domain. The spontaneous emission of this atomic medium provides a SSPP radiation with a direction satisfying Eq.~\eqref{Eq:Phase_Match}. The temporal evolution of excited atomic state~\eqref{Eq:Volterra} is a Gaussian profile with fast oscillations that appear as the absorption re-emission of the quantum plasmon is faster than the total plasmonic loss. These damped oscillations also depend on propagation time~($t_\text{SPP}$), oscillation frequency~($\omega_\text{o}$) and total Ohmic loss of the system as it is shown in Figs.~\ref{fig:two}(b) and (c). The Fourier component of this spontaneous emission is a Lorentzian line-shape with a sharp maximum in $\bm{q}\approx\bm{k}_\text{SPP}$, which guarantees this spontaneous emission propagates as directional SSPP. This oscillation regime can be controlled through our coherent loss-compensation scheme and is in agreement with previous studies on SSPP dynamics within dissipative interface~\cite{zhang2019surface}.

Our waveguide thereby acts as a high-speed single-photon switch. This switching is expected by considering excited atomic-state dynamics as $c_{3}(t)=|c_{3}(t)|\exp\{(\text{i}\omega_\text{o}-\gamma_\text{SPP})t\}$~(for $\gamma_\text{SPP}$ the total decay rate of the system). The temporal dynamics represents the strong coupling between the directional SPP and atomic medium. We also introduce resonant coupling between the directional SSPP mode and two-externally coupled plasmonic field to coherent compensation of metamaterial loss
\begin{equation}
    \text{Im}[\varepsilon_\text{N}]\approx0.2~\text{cm}^{-1},
\end{equation}
and modulate the field dispersion
\begin{equation}
    \text{Re}~(1/k_\text{N})\approx 200,
\end{equation}
yielding modified dephasing time. This SSPP dynamics is then fast, low-loss, and preserves coherence due to efficient quantum decoherence suppression~\cite{bogdanov2019overcoming}, which can be exploited to design coherent single-photon switch. Our plasmonic configuration therefore excites directional superradiant polaritonic field and consequently, the weak plasmonic field within this waveguide can amplify without need to population inversion. In \S~\ref{Sec:Surface_Plasmon_Amplification} we establish coherent amplification of the weak SPP field in the presence of this directional SSPP radiation.

\subsection{\label{Sec:Surface_Plasmon_Amplification}Surface plasmon polariton amplification}
In this section, we assume that the signal and driving laser fields are injected to the waveguide configuration through end-fire coupling technique. We consider signal as a weak pulse and driving field as a strong laser field and establish coherent amplification of weak SPP field in the presence of directional SSPP. We present this section in three steps: First, in \S~\ref{Sec:Mathematical_Formalism_Amplificaiton} we describe the general mathematical formalism towards the temporal evolution of the weak signal field. Next in \S~\ref{Sec:Dynamical_Evolution_Amplified_SPP} we elucidate the mathematical formulation of the excitation and stable propagation of SPP field within our interactive interface. And finally, in \S~\ref{Sec:Weak_Plasmonic_Amplification} we explain sufficient conditions for efficient weak field amplification, solve and plot the temporal evolution of the weak SPP field, establish coherent amplification of this plasmonic wave and investigate the amplification stability. 

\subsubsection{\label{Sec:Mathematical_Formalism_Amplificaiton}General mathematical properties of signal field dynamics}
In this section, we present mathematical formalism of our amplification scheme. Our method for quantitative description of the weak plasmonic field evolution is based on the spectral analysis of the plasmonic field in the Fourier space~\cite{PhysRevB.79.195414}, which yield Maxwell-Schr\"odinger equation for spatiotemporal domain. We achieve the spatiotemporal evolution of the weak signal field in two steps: first, we exploit the reduced Maxwell equation to describe the dynamics of the probe field and then we obtain the dynamical evolution of the dipole moment by evaluating the temporal evolution of the atomic medium.

The total plasmonic field related to the signal~($\bm{E}_\text{s}$) and driving~($\bm{E}_\text{d}$) lasers within the interaction interface is
\begin{equation}
    \bm{E}(\bm{r},t)=\sum_{m=\textrm{s,d}}\bm{E}_m(\bm{r},t)+\text{c.c.},
\end{equation}
with
\begin{equation}
    \bm{E}_m(\bm{r},t)=\mathcal{E}_l\bm{u}_l(\bm{r})\exp\{\text{i}(\bm{k}\cdot\bm{r}-\omega_lt)\},
\end{equation}
the electric field of the pumped lasers,
\begin{equation}
    \bm{u}_m(\bm{r})=c\left[k(\omega_l)\bm{e}_{z}-\text{i}k_\text{N}(\omega_l)\bm{e}_{\parallel}\right]/\varepsilon_0\omega_l,
\end{equation}
the unit signal~($\bm{u}_\text{s}$) and driving~($\bm{u}_\text{d}$) plasmonic field vectors along the interface. We define the amplitude of these fields as
\begin{equation}
    \mathcal{E}_l=\left(\frac{\hbar\omega_l}{\varepsilon_0L_{x}L_{y}L_{z}}\right),
\end{equation}
assume the interaction length along $x(y)$ direction as $L_x(L_y)$ direction and we define the field confinement factor as~\cite{Asgarnezhad_Zorgabad_2020}
\begin{equation}
    L_{z}=\sum_{j=\text{N},0}\left\{\left(\frac{\omega_l^2}{2\text{c}^2}\left[\frac{\tilde{\varepsilon}_j(|\bm{k}_j|^2+|\bm{k}|^2)}{|\bm{k}_j||\varepsilon_j^2|}\right]+\frac{\tilde{\mu}}{2|\bm{k}_j|}\right)\right\}.
    \label{Eq:Field_Confinement}
\end{equation}
Note that in defining Eq.~\eqref{Eq:Field_Confinement} we have defined
\begin{align}
    \tilde{\varepsilon}_j:=&\text{Re}\left[\frac{\partial(\omega_l\varepsilon_j)}{\partial \omega_l}\right], \; \tilde{\mu}_j:=\text{Re}\left[\frac{\partial(\omega_l\mu_j)}{\partial \omega_l}\right],
\end{align}
as effective electrical permeability and magnetic permeability of the interface, respectively. 

The driving and signal fields are coupled to the atomic medium with Rabi frequencies $\Omega_\text{d}$ and ~$\Omega_\text{s}$ are tightly confined to the interface by transversely evanescence coupling functions $\zeta_\text{d}(z)$ and $\zeta_\text{s}(z)$ respectively~(this coupling function is achieved in Refs.~\cite{Asgarnezhad_Zorgabad_2020,PhysRevA.98.013825} and we present the mathematical steps towards coupling function derivation in appendix~\ref{Maxwell_Schrodinger_Amplification_SPP}). Specifically, we employ mapping
\begin{equation}
    \Omega_{m}:=\zeta_{m}(z)\Omega_{m}; m\in\{\text{d},\text{s}\},
\end{equation}
and consider the effect of field confinement by exploiting field averaging
\begin{equation}
    \braket{\mathcal{F}(z)}=\frac{\int_{-\infty}^{+\infty}\text{d}z\;\zeta^*(z)\mathcal{F}(z)}{\int_{-\infty}^{+\infty}\text{d}z\;|\mathcal{F}(z)|^2}.
\end{equation}
In what follows, we assume the signal field to be weak and obtain its dynamics in the presence of directional superradiant emission.

We also assume this weak signal field as a plasmonic plane wave with constant phase
\begin{equation}
    \varphi:=\tilde{\omega}_\text{SPP}(t-x/v_\text{SPP}),
\end{equation}
and with wavenumber $k_\text{SPP}$ that couples $\ket{1}\leftrightarrow\ket{3}$ atomic transition and propagates along the interaction interface with group velocity
\begin{equation}
    \bm{v}_\text{SPP}=\left[\frac{\partial\bm{k}_\text{SPP}}{\partial\omega}\right]^{-1}.
\end{equation}
Our quantitative description of the signal SPP wave dynamics, is based on Fourier optics, which is formulated in \S~\ref{Sec:Fourier_Optics}. 

We consider a Fourier space that is characterized with a real wavevector $\bm{q}$ and complex $\tilde{\omega}$. Moreover we choose Eq.~\eqref{Eq:Green_Funtion_Interface} as Dyadic green tensor of the interaction interface and consider the weak signal field Rabi frequency within spectral domain as
\begin{equation}
    \tilde{\Omega}_\text{s}\approx\int\;\frac{\text{d}^2\bm{q}}{(2\pi)^2}\bm{\tilde{\mathcal{A}}}_\text{SPP}\hat{j}\text{e}^{\text{i}[(\bm{q}-\bm{k}_\text{SPP})\cdot\bm{r}-\tilde{\omega} t]}.
    \label{Eq:Fourier Dynamics}
\end{equation}
We note that our hybrid interface is highly dissipative and we achieve ultra-low loss operation only for small frequency deviation of $\omega_\text{SPP}$. This SPP field is then highly unstable and lossy for
\begin{equation}
    \bm{q}\gg\bm{k}_\text{SPP},\;\;\;\; \omega\gg\omega_\text{SPP},
\end{equation}
for a small frequency and wavenumber deviations from $\omega_\text{SPP}$ and $\bm{k}_\text{SPP}$ as NIMM layer and we achieve stable propagation for small wavevector~($\delta\bm{k}:=\bm{q}-\bm{k}_\text{SPP}$) and frequency~($\delta\omega$) deviations
\begin{align}
    \bm{q}=&\bm{k}_\text{SPP}+\mathcal{O}(\bm{q}-\bm{k}_\text{SPP}),\\
    \omega=&\omega_\text{SPP}+v_\text{SPP}\delta\omega+\mathcal{O}(\delta\omega^{2}).
\end{align}
Therefore we treat these quantities as perturbation parameters~\cite{PhysRevA.98.013825,PhysRevA.99.051802}. Consequently, the weak SPP stably propagate within the interface due to ultra-low Ohmic loss and suppressed nonlinearities~\cite{PhysRevA.81.033839,PhysRevA.85.050303}.

Finally, we achieve the reduced Maxwell equation for this weak signal field within our characteristic $\tilde{\omega}$-$\bm{q}$ space by considering the stable propagation regime, taking derivatives with respect to $t$ and $\bm{r}$ as
\begin{equation}
    \left(\frac{\partial}{\partial t}+\bm{v}_\text{SPP}\cdot\bm{\nabla}\right)\Omega_\text{s}=\text{i}\mathcal{C}\tilde{\rho}_{31},
   \label{Eq:Reduced_Maxwell_Equation}
\end{equation}
with
\begin{equation}
    \mathcal{C}=N_\text{a}\frac{\pi\mathcal{A}_\text{SPP}}{\gamma_\text{SPP}}.
\end{equation}
Note that our Eq.~\eqref{Eq:Reduced_Maxwell_Equation} is similar to Maxwell-Schr\"odinger equations obtained in earlier works~\cite{PhysRevX.3.041001}, however~\eqref{Eq:Reduced_Maxwell_Equation} differs from previous works due to incorporating dissipation and dispersion of the surface-plasmon mode to the atomic medium evolution. We note that the dipole moment of the system would be proportional to the $\ket{3}\leftrightarrow\ket{1}$ atomic transition which is characterized in Eq.~\eqref{Eq:Reduced_Maxwell_Equation} by $\tilde{\rho}_{31}$. This term in \eqref{Eq:Reduced_Maxwell_Equation} is atomic coherence term, which is related to atomic transition amplitudes as
\begin{equation}
    \tilde{\rho}_{31}=\left[c_{3}^{*}(t)c_{1}(t)+c_{3}^{*}(t)c_{1}(t)\right]\text{e}^{-\text{i}\varphi-\gamma_\text{SPP}t}.
    \label{Eq:Atomic_Dipole_Final}
\end{equation}
Eq.~\eqref{Eq:Reduced_Maxwell_Equation} commensurate with coherence term \eqref{Eq:Atomic_Dipole_Final} describes the dynamics of the weak signal field within this dissipative interface. In the next section we discuss the amplification condition and establish that this weak plasmonic field can be amplified for specific modulation of the driving field.

\subsubsection{\label{Sec:Dynamical_Evolution_Amplified_SPP} Spatiotemporal evolution of weak plasmonic field: Quantitative description}
In this section, we describe the dynamical evolution of the weak plasmonic field and discuss the opportunities to achieve the spatiotemporal dynamics of this signal polaritonic wave. To stable propagation of weak signal field, we suggest a strong driving field with Rabi frequency $\Omega_\text{d}$ that is orthogonally polarized, its corresponding Rabi frequency $\Omega_\text{d}^{(1)}$ couple an intermediate state $\ket{a}$ to atomic transitions through $\ket{1}\leftrightarrow\ket{a}$ and we assume $\Omega_\text{d}^{(2)}$ drives $\ket{1}\leftrightarrow\ket{3}$ transition for different polarizations~(see Fig.~\ref{fig:Figm}). We achieve the coherent term \eqref{Eq:Atomic_Dipole_Final} for this excited atomic medium by employing the Schr\"odinger approach. First we obtain the dynamics of weak signal plasmonic field in a general case and then we describe sufficient conditions for which an efficient weak field amplification can be achieve. 

\emph{General description}- Here, we evaluate temporal dynamics of the plasmonic field within our interaction interface, for weak signal and orthogonally polarized strong driving lasers. We assume a general case for which the weak signal field is linearly polarized and the strong field is modulated as a circularly polarized field
\begin{align}
    \Omega_\text{d}(x,t)=\bar{\Omega}_\text{d}^{(1)}(x,t)\hat{\epsilon}_{+}+\bar{\Omega}_\text{d}^{(2)}(x,t)\hat{\epsilon}_{-},
    \label{Eq:Deriving field pattern}
\end{align}
for right~($\epsilon_+$), ($\epsilon_{-}$) circular polarization and 
\begin{equation}
    \epsilon_{\pm}=(\bm{e}_{x}\pm\text{i}\bm{e}_{y})/\sqrt{2}.
\end{equation}
The Hamiltonian of the system in the presence of these two modulated fields is
\begin{align}
    H_\text{I}=&\hbar\bigg[\zeta_\text{d}^{(1)}(z)\Omega_\text{d}^{(1)}\text{e}^{-\text{i}\omega_{31}t}\ket{3}\bra{1}+\zeta_\text{d}^{(2)}(z)\Omega_\text{d}^{(2)}\text{e}^{-\text{i}\omega_{a1}t}\nonumber\\&\times\ket{a}\bra{1}+\zeta_\text{d}^{(1)}(z)\Omega_\text{s}\text{e}^{-\text{i}\omega_{31}t}\ket{3}\bra{1}+\text{c}.\text{c}.\bigg].
\end{align}
for $\Omega_{m}$ the Rabi frequency of the signal and deriving fields
\begin{align}
    \Omega_\text{s}=&\frac{E_\text{s}d_{31}}{\hbar},\\
    \Omega_\text{d}^{(1)}=&\frac{E_\text{d}d_{31}}{\hbar},\; \Omega_\text{d}^{(2)}=\frac{E_\text{d}d_{a1}}{\hbar},
\end{align}
we consider the relaxation rate of the $\ket{1}$~($\ket{3}$) as $\gamma_{1}$~($\gamma_{3}$), respectively, and assume the relaxation rate of the intermediate state as $\gamma_\text{a}$. Next, we assume the following Ansatz for atomic transition amplitudes
\begin{equation}
    \ket{\psi(t)}=c_{1}(t)\ket{1}+c_{a}(t)\ket{a}+c_{3}(t)\ket{3}.
\end{equation}
To achieve the dynamical evolution of the atomic states we assume simplification for field confinement
\begin{equation}
    \zeta_\text{s}(z)\approx\zeta_\text{d}(z):=\zeta(z),
\end{equation}
employ mapping to deriving fields
\begin{equation}
    \braket{\zeta(z)\Omega_l}\mapsto\Omega_l,
\end{equation}
and consider the rotated atomic transitions
\begin{equation}
    c_\text{3}\text{e}^{-\text{i}\omega_{31}t}\mapsto\tilde{c}_{3},\; c_\text{1}\text{e}^{-\text{i}\omega_{\text{a},1}t}\mapsto\tilde{c}_{1}.
\end{equation}
Consequently, we evaluate the atomic state evolution as
\begin{align}
    \dot{c}_{1}(t)+\text{i}\Omega_\text{d}^{(1)}c_{3}(t)+\text{i}\Omega_\text{d}^{(2)}c_{a}(t)-\frac{\gamma_\text{3}}{2}c_{3}-\frac{\gamma_\text{a}}{2}c_{a}=&0,\label{Eq:Ground_State_Total}\\
    \dot{c}_{3}(t)+\left(\text{i}\omega_{31}+\frac{\gamma_3}{2}\right)c_{3}(t)+\text{i}\Omega_\text{d}^{(1)}c_{3}+\text{i}\Omega_\text{d}^{(2)}c_{1}=&0.\label{Eq:Excited_State_Total}
\end{align}
Finally we introduce signal~($\Delta_\text{s}=\omega_\text{s}-\omega_{31}$) and driving field detunings~($\Delta_\text{d}=\omega_\text{s}-\omega_{31}$) to achieve the temporal evolution of the atomic states. 

We then achieve the most general form of the coupled Maxwell-Schr\"odinger equation for signal field by taking the time derivatives of Eq.~\eqref{Eq:Reduced_Maxwell_Equation} commensurate with temporal evolution of coherence term $\dot{\tilde{\rho}}_{31}$, which is achieved by taking derivative with respect to time of Eq.~\eqref{Eq:Atomic_Dipole_Final} and substituting $\dot{c}_{j}(t)$ from Eqs.~\eqref{Eq:Ground_State_Total} and~\eqref{Eq:Excited_State_Total}. The Rabi frequency of this weak plasmonic signal then followed by the following partial differential equation
 \begin{equation}
   \left(\frac{\partial}{\partial t}+\beta(x,t)\right)\left(\frac{\partial}{\partial t}+\bm{v}_\text{SPP}\cdot\bm{\nabla}\right)\Omega_\text{s}=\text{i}\mathcal{C} f(x,t)\Omega_\text{s},
   \label{Eq:Main equaiton}
\end{equation}
here the coefficients $\beta$ and $f$ are
\begin{align}
    \beta:=&\tilde{\Delta}+\frac{\bar{\Omega}_\text{d}^{(1)2}+\bar{\Omega}_\text{d}^{(2)2}}{-\text{i}\omega_\text{SPP}+\gamma_\text{SPP}}+\frac{4\bar{\Omega}_\text{d}^{(1)}\bar{\Omega}_\text{d}^{(2)*}\cos(2\varphi)}{\omega_\text{SPP}+\text{i}\gamma_\text{SPP}},\nonumber\\
    f:=&1-\frac{2\bar{\Omega}_\text{d}^{(1)2}}{\tilde{\Delta}(-\text{i}\omega_\text{SPP}+\gamma_\text{SPP})}-\frac{2\bar{\Omega}_\text{d}^{(1)2}+\bar{\Omega}_\text{d}^{(2)2}\exp\{2\text{i}\varphi\}}{\omega_\text{SPP}(-\text{i}\omega_\text{SPP}+\gamma_\text{SPP})}.
\end{align}
We express the mathematical steps towards Eq.~\eqref{Eq:Main equaiton} in appendix~\ref{Sec:Surface_plasmon_amplification_Appendix}. Using Eq.~\eqref{Eq:Main equaiton} we demonstrate coherent amplification by solving and plotting the dynamics of the Rabi frequency of the $\Omega_\text{s}$, and next we explore its dependence on system parameters to obtain a sufficient conditions for exciting intense plasmonic waves.

\emph{Assumptions and feasibility in experiment}- Now we present our assumptions to achieve spatiotemporal dynamics of the weak plasmonic field in the presence of polaritonic superradiant emission and next we give realistic parameters to test the feasibility of the scheme. Here, the atomic medium are coupled to a weak signal and orthogonally polarized strong driving fields, hence the atomic transition amplitude would be affect by these injected fields. Consequently, $c_{\imath,\text{s}}$; $\imath\in\{1,\text{a},3\}$ represents the $\ket{\imath}$ atomic states affect by signal field and $c_{\imath,\text{d}}$ describes its evolution for driving field. We achieve dynamical evolution of the weak field \eqref{Eq:Main equaiton} and amplification by assuming modifications to atomic transition amplitudes, coherent term and Rabi frequency as
\begin{align}
    c_{\imath}=&c_{\imath,\text{s}}+c_{\imath,\text{d}},\\
    \tilde{\rho}_{31}=&\tilde{\rho}_{31,\text{s}}+\tilde{\rho}_{31,\text{d}},\\
    \Omega(\bm{r},t)=&\Omega_\text{s}(\bm{r},t)+\Omega_\text{d}(\bm{r},t).
\end{align}
We consider the modulations due to weak signal field as
\begin{align}
  \Omega_\text{s} & =\zeta(z)\Omega_\text{s}(t)\exp\{\text{i}\varphi\}, \\ c_{\imath,\text{s}} & =\zeta(z)c_{\imath,\text{s}}(t)\exp\{\text{i}\varphi\}+\text{c}.\text{c}.,
\end{align}
and employ Fourier analysis, characterized by real wavevector~($\bm{q}$) and complex perturbation frequency~($\tilde{\omega}$). 

The weak signal plasmonic field then experience amplification and Eq.~\eqref{Eq:Main equaiton} describes its spatiotemporal dynamics. In obtaining this equation, we neglect the temporal evolution of the ground state
\begin{equation}
    \dot{c}_{1,s}\ll1,
\end{equation}
assume copropagating electric signal and driving fields $\bm{E}=\bm{E}_\text{s}+\bm{E}_\text{d}$, we consider the phase of the plasmonic field~($\phi=\bm{k}_\text{d}\cdot\bm{r}-\omega t$) to be constant, and we take into account the plasmonic evanescence coupling by employing field averaging technique~(i.e. $\braket{\zeta(z)\Omega_l}\mapsto\Omega_l$~\cite{PhysRevA.98.013825}). Here, we test the feasibility of our scheme by employing the realistic parameters. For atomic medium $\omega_{21}=2.54~\text{eV}$ and $\omega_{31}\approx2.04~\text{eV}$~\cite{utikal2014spectroscopic,eichhammer2015spectroscopic}. We set $I_\text{p}=30~\mu\text{W}$ and propagation length as $L=200~\mu\text{m}$. To describe NIMM layer we use Drude-Lorentz model with $\varepsilon_{\infty}=\mu_{\infty}=1.2$, $\omega_\text{e}=1.37\times10^{16}~\text{s}^{-1}$, $\omega_\text{m}=10^{15}~\text{s}^{-1}$, $\gamma_\text{e}=2.37\times10^{13}~\text{s}^{-1}$ and $\gamma_\text{m}=10^{12}~\text{s}^{-1}$~\cite{shalaev2007optical}. Then for $\omega_{31}$ frequency transition $\text{Im}[\mathcal{A}(z,z_\text{at};k_\text{SPP})]\approx3.2\times 10^{10}$ and $v_\text{SPP}\approx2.61\times10^{-2}\text{c}$. We employ Eqs.~\eqref{Eq:Volterra} and \eqref{Eq:Kernel_Volterra} commensurate with these realistic parameters to achieve the SSPP dynamics.

\subsubsection{\label{Sec:Weak_Plasmonic_Amplification} Weak plasmonic field amplification and stability analysis}
In this section, we obtain sufficient conditions for weak polaritonic amplification within our hybrid plasmonic interface. Although this amplification is achieved for various system parameters, we assume specific modulation for coupling field intensities and detunings for efficient propagation of the signal plasmonic field and discuss the amplification stability. Now, we present the results for the case that a linearly polarized weak signal and orthogonally polarized strong driving fields are injected to our interaction interface.

We employ slowly varying amplitude approximation in our analysis~\cite{boyd2003nonlinear}, assume $\gamma_1\approx0$, and we take $\gamma_\text{SPP}$ to be a perturbation parameter due to the controllability of the virtual gain by external laser fields. Therefore, solving~\eqref{Eq:Main equaiton} commensurate with the Schr\"odinger equation for driving~($\Omega_\text{d}^{(1)}\approx\Omega_\text{d}^{(2)}:=\Omega_\text{d}$) and weak signal plasmonic fields and keeping the terms up to $\gamma_\text{SPP}$ and $\omega_\text{SPP}$, yields
\begin{equation}
    \delta\omega_\text{SPP}\approx\tilde{\Delta}+\frac{|\Omega_\text{d}|^2}{\omega_\text{SPP}}\left[2+\text{i}\frac{\gamma_\text{SPP}}{\omega_\text{SPP}}\left(1-\frac{|\Omega_\text{d}|}{\omega_\text{SPP}}\right)\right].
    \label{Eq:Frequency_Shift}
\end{equation}
Note that all the terms within \eqref{Eq:Frequency_Shift} are frequency shifts. The additional term, which depends on the driving field amplitude, would corresponds to Stark shift~\cite{Cho:16} that limits the amplification efficiency in the experiment. This unwanted effect would also reduce the amplification efficiency in our system. First we note that leading term in Eq.~\eqref{Eq:Frequency_Shift} depends on driving field intensity that we refer as Stark shift~\cite{PhysRevX.3.041001}. The effect perturbs the excitation frequency of the SPP field due to spectral broadening, thus the SPP field frequency deviates from characteristic frequencies for which amplification can be achieved~($\omega_\text{SPP}\neq\omega_\text{ch}$). The Stark shift then destroys parametric resonance, suppresses the gain and consequently reduces the amplification efficiency.

In order to reduce this effect and efficient amplification, we suggest the driving field amplitude modifies within the interaction plane ($x$-$y$) as
\begin{align}
    \Omega_\text{d}(x,t)=\bar{\Omega}_\text{d}^{(1)}\text{e}^{\text{i}(\bm{k}_\text{d}\cdot\bm{r}-\omega_\text{d}t)}\hat{\epsilon}_{+}+\bar{\Omega}_\text{d}^{(2)}\text{e}^{-\text{i}(\bm{k}_\text{d}\cdot\bm{r}-\omega_\text{d}t)}\hat{\epsilon}_{-},
    \label{Eq:Deriving_field_pattern}
\end{align} 
and its wavenumber modulates as
\begin{equation}
    \bm{k}_\text{d}\approx \bm{k}_\text{SPP}+\delta\bm{k},
\end{equation}
for $\delta\bm{k}\ll\bm{k}_\text{SPP}$. The resonant interaction between these weakly modulated plasmonic fields and directional SSPP preserves quantum coherence, reduce the unwanted Stark shift, and thereby provides efficient spectral component. In this case, Eq.~\eqref{Eq:Main equaiton} also changes to the Mathieu's differential equation that supports the amplification of the weak signal field within characteristic dispersion curves~\cite{nayfeh2008perturbation}.  
\begin{figure}
    \includegraphics[width=1\columnwidth]{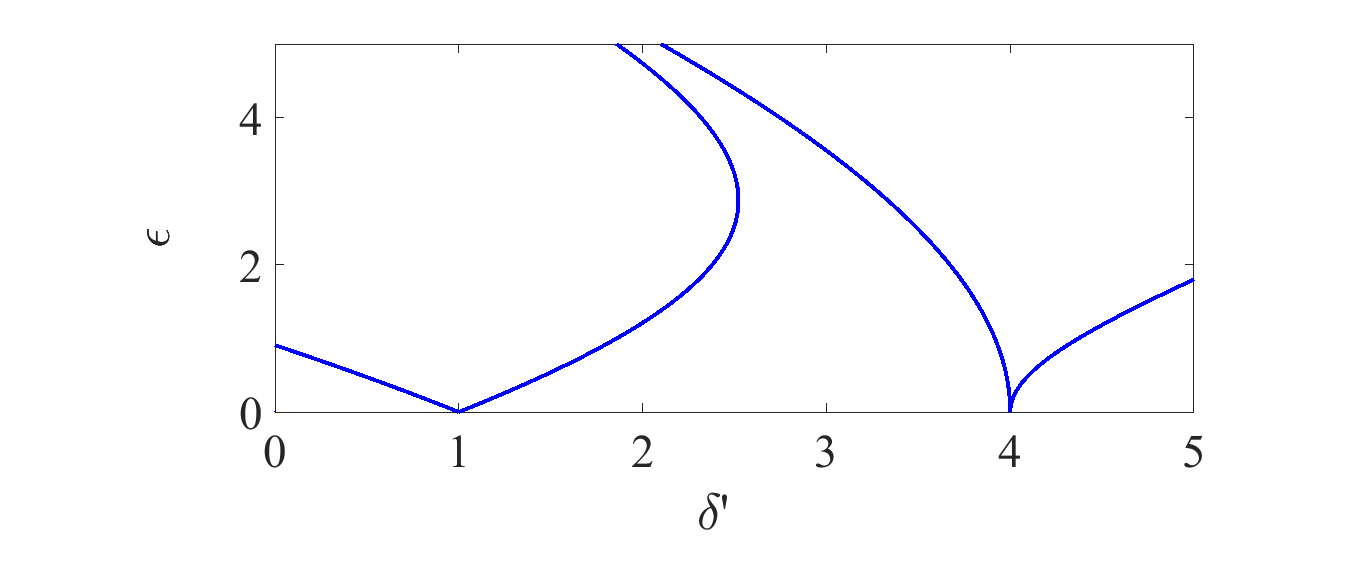}
    \caption{Stability diagram of the amplified SPP field: The presented curves in $\epsilon-\delta'$ plane denote the efficient amplification. Simulation parameters are $\beta=0.001$, $\omega_\text{d}=2\pi\times20~\text{MHz}$, $m=1$, and we assume $\bm{\nabla}\Omega_\text{s}=0$, $|\Omega_\text{d}|^2=10~\text{MHz}$, $\omega_\text{SPP}=\omega_{31}$ and $\gamma_\text{SPP}=0.004~\text{s}^{-1}$. Other parameters are the same as Fig.~\ref{fig:two}.}
    \label{fig:three}
\end{figure}

Specifically, we achieve the coherent amplification and investigate its stability within the atomic medium-NIMM interface in the case in the presence of directional plasmonic superradiant excitation \eqref{timed-dicke state} within our interaction interface. For our proposed realistic parameters, $\delta\bm{k}_\text{d}\ll1$ and by taking into account~\eqref{Eq:Deriving_field_pattern}, we achieve
\begin{equation}
    \tilde{\rho}_{11}-\tilde{\rho}_{33}\approx[1+\eta\cos(2\delta t)],
    \label{Eq:Plasmonic_Noise}
\end{equation}
$\eta$ the modulation parameter that depends on the plasmonic noise~\cite{Asgarnezhad-Zorgabad:20}. Now we define $\delta'$ as normalized detuning, and $\epsilon$ the normalized amplitude that depends on system parameters. We assume these quantities as our control parameters and investigate the amplification efficiency for this tunable system parameters. In this case, Eq.~\eqref{Eq:Main equaiton}, which represents our simplified Maxwell-Schr\"odinger equation then reduces to Mathieu-like equation
\begin{equation}
    \ddot{\Omega}_\text{s}+\delta'[1+\epsilon\cos(2\delta t)]\Omega_\text{s}=0,
    \label{Eq:Matuie_Equation}
\end{equation}
for $\tilde{\Delta}\ll1$, and
\begin{align}
    \delta'=\mathcal{C}f(x,t)+v_\text{g}^{2}k_\text{SPP}^{2}/4,\label{Eq:Matheiu_Parameter_delta}\\
    \epsilon=\frac{\mathcal{C}f(x,t)\eta}{\mathcal{C}f(x,t)+v_{g}^{2}k_\text{SPP}^{2}/4}.
    \label{Eq:Matheiu_Parameter_epsilon}
\end{align}
To achieve weak plasmonic field amplification without population inversion, we solve \eqref{Eq:Matuie_Equation} for weak plasmonic field modulation. Here we consider the initial plasmonic amplitude as $\Omega_\text{s}^{(0)}=0.5~\text{MHz}$ and introduce noise as $|\eta|=0.08\Omega_\text{s}^{(0)}$, and we finally assume $\dot{\Omega}_\text{s}(t=0)=0$~\cite{Asgarnezhad-Zorgabad:20}.

As we establish in \S~\ref{Sec:Mathematical_Formalism_Amplificaiton} and \S~\ref{Sec:Dynamical_Evolution_Amplified_SPP} this Mathieu equation \eqref{Eq:Matuie_Equation} is achieved by applying the modulated polarized driving field and adjusting system parameters to Eq.~\eqref{Eq:Main equaiton}. The parameters that describe the Mathieu equation, would consequently relate to the parameters describing \eqref{Eq:Main equaiton} that depend on coupling lasers and waveguide parameters. Moreover, the parameters in \eqref{Eq:Matuie_Equation} would also depend on weak signal field injection,that depends on plasmonic noise. This noise will affect the population difference due to \eqref{Eq:Plasmonic_Noise} and consequently affects the Mathieu equation parameters. We express these parameters in terms of system parameters in Eqs.~\eqref{Eq:Matheiu_Parameter_delta} and \eqref{Eq:Matheiu_Parameter_epsilon}.  

We then achieve the amplification based on two different mechanisms. 
\begin{itemize}
    \item Exploiting multiple-scale time variable techniques~\cite{nayfeh2008perturbation}. We perturbed the time as 
    \begin{equation}
        t_{l}=\varepsilon^{l}t,\;\;\; l\in\{0,1\},
    \end{equation} a
    and we assume the asymptotic expansion
    \begin{equation}
        \Omega_\text{s}=\Omega_\text{s}^{(0)}+\mathcal{O}(\Omega_\text{s}^{(1)}).
    \end{equation}
    The zeroth-order Ansatz $\Omega_\text{s}^{(0)}\sim\exp\{\delta't\}$ amplifies for specific oscillation frequency $\omega_0$, which is different from oscillation frequency of quantum SPP mode $\omega_\text{o}$, if 
\begin{equation}
    \delta':=\omega_\text{0}^2,\;\delta\approx\omega_\text{0}/m,
\end{equation}
with $m=1,2,3,\cdots$. Here, the nonlinear processes generate frequency combs~($\delta\omega_\text{comb}\approx10~\text{MHz}$~\cite{PhysRevA.99.051802}) that leakage the SPP energy to other spectral component, and reduce amplification efficiency. Consequently, the frequency grid should be small compared to this frequency comb spacing.
\item Using the gain modulation. To obtain the gain threshold we set
\begin{align}
    \delta'=&\delta'^{(0)}+\varepsilon\delta'^{(1)}+\mathcal{O}(\delta'^{(2)}),\\
    \Omega_\text{s}=&\Omega_\text{s}^{(0)}+\varepsilon\Omega_\text{s}^{(1)}+\mathcal{O}(\Omega_\text{s}^{(2)}),
\end{align}
consider the zeroth order solution as
\begin{equation}
    \Omega_\text{s}^{(0)}\sim A\text{e}^{\text{i}t_0},   
\end{equation}
and first order perturbation as
\begin{equation}
    \Omega_\text{s}^{(1)}:=A(\delta')\exp\{\delta t\},
\end{equation}
for
\begin{equation}
    A(\delta'):=\text{Re}[A(\delta')]+\text{i}\text{Im}[A(\delta')],
\end{equation}
the amplitude of the superradiant mode under weak perturbation limit. Plugging into Eq.~\eqref{Eq:Matuie_Equation} we obtain the gain threshold for amplification of the weak SPP wave as 
\begin{equation}
    2\gamma_1\approx\sqrt{1-4\delta'^{(1)^2}}.
    \label{Eq:Gain_mdulation}
\end{equation}
The plasmonic system provides gain for the detuning frequency characterized by \eqref{Eq:Gain_mdulation}. As high frequency deviation produces gain at least for two specific frequencies, which reduces the amplification efficiency, the frequency deviation should be narrow to produce gain only for $\omega_0$.
\end{itemize}

\begin{figure}
    \centering
    \includegraphics[width=1\columnwidth]{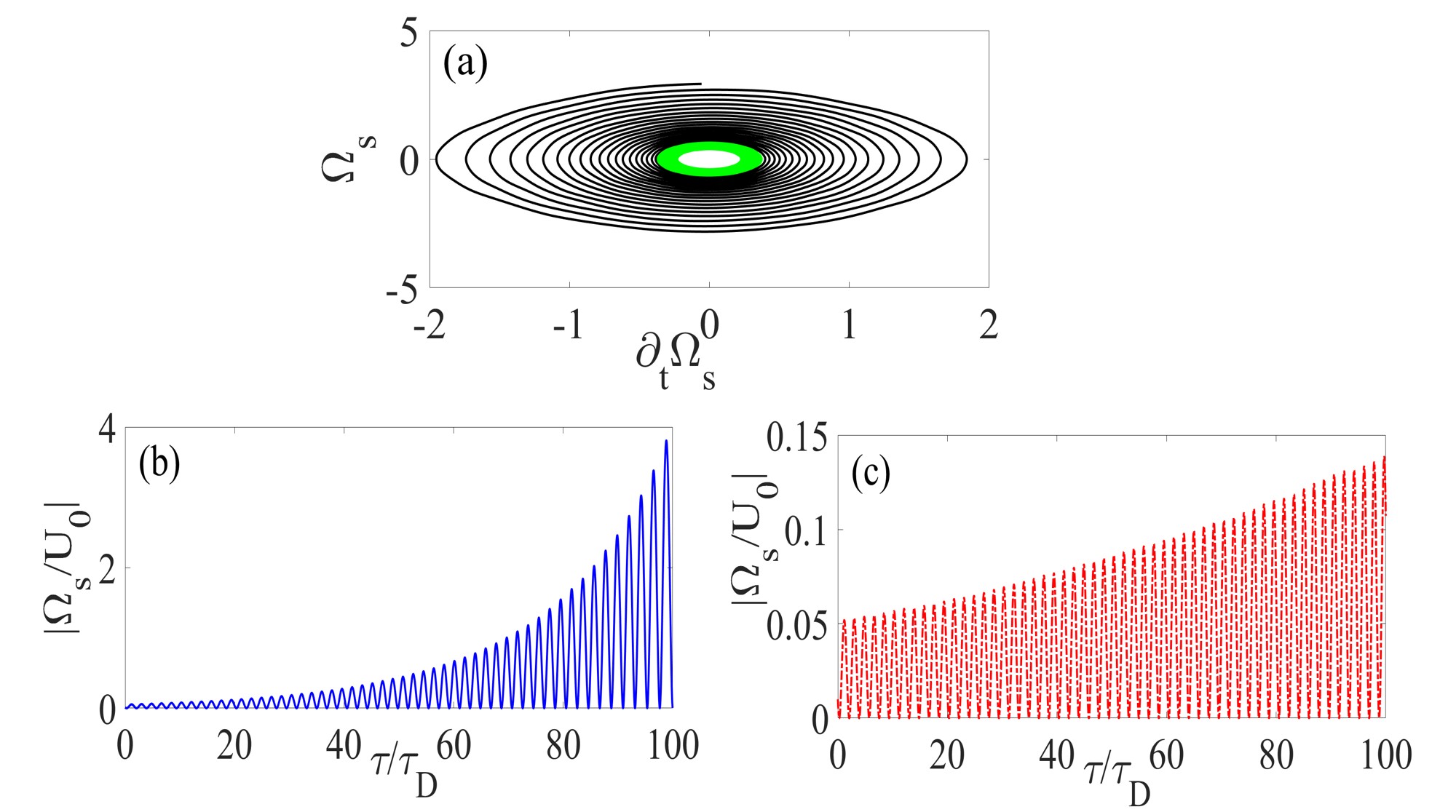}
    \caption{Dynamics of the SSPP: The figure represents the superradiant emission with efficient amplification condition Eq.~(\ref{Eq:Best Amplification})~(blue solid line) and without considering Eq.~(\ref{Eq:Best Amplification}). The parameters for these plots are $|\Omega_\text{d}^{(0)}|=10~\text{MHz}$~(for red line), $\delta'^{(0)}=0$, $\delta'^{(1)}=\omega_0$. Other parameters are the same as Figs.~\ref{fig:two} and \ref{fig:four}.}
    \label{fig:four}
\end{figure}

The stable amplification for sufficiently large $\delta'-\epsilon$ is induced to this chaotic plasmonic interface through resonant interaction between weak plasmonic field and SSPP emission~\cite{ruby1996applications}. In this case, \eqref{Eq:Matuie_Equation} is a homogeneous equation that yields stable amplification of the weak plasmonic field within certain spectral region represented in Fig.~\ref{fig:three}. Specifically, our simulation demonstrates that the efficient SPP amplification with $\epsilon\neq0$ can be achieved for $1<\delta'<4$. We extend our model to the spatially dependent amplification of the SSPP by considering ($\bm{v}_\text{SPP}\cdot\bm{\nabla}$) as a correction term and treating $\beta$ as a perturbation term. The weak plasmonic field is modulated as
\begin{equation}
    \Omega_\text{s}:=\Omega_{s}^{(0)}[1+p(x)]\exp\{\delta t+\text{i}\mathcal{K}x\},
\end{equation}
with $\mathcal{K}:=|\bm{k}_\text{SPP}|$. Now we substitute this equation into Eq.~\eqref{Eq:Reduced_Maxwell_Equation} and and employ linear stability analysis~\cite{PhysRevA.85.023822}. In this case, the nonlinearity would be reduced and this SPP wave amplifies within the spectral stability diagram represented in Fig.~\ref{fig:three} if the system parameters are modulated as
\begin{equation}
    \frac{4\Omega_\text{d}^{2}}{\omega_\text{SPP}+\text{i}\gamma_\text{SPP}}+\text{i}\mathcal{K}|\bm{v}_\text{SPP}|\approx0.
    \label{Eq:Best Amplification}
\end{equation}
 The amplification pattern corresponds to this propagated plasmonic mode is represented in Fig.~\ref{fig:four}. The robustness of amplification depend on low SPP field propagation and robust directional SSPP launching. Due to loss-compensation scheme, we expect the stable propagation of SPP field and hence robust amplification of SPP field is produced only for robust SSPP, which is achieved as waveguide decay rate~($\Gamma_\text{W}$) exceed the free-space one~($\Gamma_\text{F}$)~\cite{PhysRevA.99.043807}. For our hybrid waveguide the decay rate is controllable, $\Gamma_\text{W}=5\times10^{-3}\gamma_\text{F}$, consequently much photon collection and robust SSPP are expected with error scale up to $\varepsilon_\text{er}\approx0.12$, which demonstrate the robustness of our scheme. 

\subsection{\label{Sec:Detection_Amplified}Detection of amplified plasmonic field}
This amplified plasmonic wave can be detected in an experiment. We underpin our method for measuring the amplified SPP field, based on the far-field pattern of the propagated plasmonic waves. We define this far-field radiation $\bm{E}_\text{super}$ in terms of Poynting vector as
\begin{equation}
    \bm{S}=\frac{1}{2}\text{n}_\text{eff}\varepsilon_{0}|\bm{E}(\bm{r},t)|^{2}\bm{e}_{\parallel},
\end{equation}
for $\text{n}_\text{eff}$ the effective refractive index of the interface and $\bm{e}_\parallel$ the unit vector of the superadiant field along the interface. This electric field is confined to the interface with an effective thickness $z=z_\text{at}$, and we evaluate this plasmonic field for our hybrid interface in appendix~\ref{Sec:Calculation_far_Field_Superradiant}. The intensity profile of this signal plasmonic wave thereby depends on the interference pattern of this far-field emission, which is constructive for azimuth angle $\phi$ and polar angle $\theta$ satisfying 
\begin{equation}
    \mathcal{K}\sin\theta(\cos\phi x+\sin\phi y)=2m\pi.
    \label{Eq:Far_Field}
\end{equation}
For measuring the amplified plasmonic wave intensity, we suggest a detection system that is placed in a spatial coordinate obtained by solving Eqs.~\eqref{Eq:Phase_Match} and~\eqref{Eq:Far_Field}~(we represent the detailed derivation of the Eq.~\eqref{Eq:Far_Field} in~\ref{Sec:Calculation_far_Field_Superradiant}). We detect this wave for a certain deviations in detuning, amplitude and frequencies. This effect is detectable for deviation in frequency detunings that introduce negligible atomic absorption and provide loss-compensation. Also, for signal amplitudes with suppressed intensity dependent nonlinear effects such as Kerr nonlinearity and higher-order dispersion, an amplification with coherent spectral properties is expected. Finally, only those frequency deviations can be efficiently scattered and detected via our plasmonic configuration that are resonant with our characteristic Bragg frequencies.

\section{\label{Sec:Discussion}Discussion}
We begin this section by summarizing our method of the weak plasmonic field amplification without need to population inversion at the interface between atomic medium and NIMM layer. We note that our quantitative description for weak plasmonic amplification is based on three main steps, namely, (i) exciting directional superradiant emission, (ii) injecting weak signal and strong deriving fields, and evaluating the spatiotemporal evolution of the weak signal field at interaction interface, and (iii) characterizing sufficient condition for amplifying weak plasmonic field. In this work we provide mathematical formalism of SSPP in \S~\ref{Sec:Mathematical_Formalism} and qualitative description in \S~\ref{Sec:Dynamical_Evolution_SSPP}. Using this SSPP we achieve the dynamics of SPP waves for injecting weak signal and strong driving lasers in \S~\ref{Sec:Mathematical_Formalism_Amplificaiton} and represent spatiotemporal dynamics in \S~\ref{Sec:Dynamical_Evolution_Amplified_SPP}, and finally we present conditions for weak SPP amplification in \S~\ref{Sec:Weak_Plasmonic_Amplification}. Consequently, our steps for amplifying weak SPP field are qualitatively and quantitatively connected.

As we establish in Sec.~\ref{Sec:Results}, the weak SPP wave amplifies within the atomic medium-NIMM layer interface through resonant coupling between this plasmonic mode and directional superradiant SPP. Our method takes advantage of the constructive interference between two contra-propagating plasmonic modes to suppress the Ohmic loss, and we achieve stability by coupling the atomic ensemble dynamics of injected signal plasmonic waves.

This quantum-plasmonic configuration, thereby serves as an efficient plasmonic transistor~\cite{Dzedolik:19}. The operation of this device is based on the generation of controllable output plasmonic fields for a weak input SPP signal. Using the realistic parameters for the atomic medium and for the NIMM layer, and by controlling the external laser-field intensities, the Ohmic loss and consequently the output power of the SPP field can be coherently controlled. Therefore, our scheme acts as a field-effect plasmonic transistor with a controllable fast-switch that operates in the optical frequency range. 

Finally, our proposed apparatus also serves as a surface plasmon laser. We explain the lasing operation base on three effects: (i) this scheme preserves coherence due to quantum decoherence suppression and exploiting coherent loss-compensation mechanism, (ii) we achieve a strong directionality for the amplified SPP wave by resonant interaction between directional superradiant SPP emission and weak plasmonic field, and (iii) our amplification scheme prevents the generation of amplified spontaneous emission of the plasmonic wave~\cite{meng2013wavelength} due to weak-field seeding and the spectral width of this intense emission is much narrower compared to other schemes due to limitations induced by the collective oscillation of the atomic medium.

\section{\label{Sec:Conclusion}Conclusion}
In summary, we devise a quantum-plasmonic waveguide, that exploits superradiant emission of radiation to produce coherent intense SPP waves. Our scheme incorporates source-waveguide-detection triplet and the waveguide~(as a sub-element of our apparatus) is a hybrid structure comprises atomic medium doped in a transparent dielectric situated above a NIMM layer. We establish SSPP dynamics based on non-Markovian spontaneous emission of atomic ensemble, which is achieved by coupling quantum plasmonic mode to collective atomic oscillation. Our framework for spatiotemporal dynamics of the weak plasmonic field within the interaction interface is based on Fourier component evolution, yielding coupled Maxwell-Sch\"odinger like equation.  We employ a coherent loss-compensation scheme and establish a resonant interaction between SSPP mode and signal plasmonic field to suppress the quantum decoherence of the amplified directional SPP wave. The quantum gain for this weak seeded plasmonic field is produced by introducing parametric resonance between SSPP and modulated stable plasmonic field. Our amplification scheme is efficient and robust against the photon loss and this amplification is only achieved for specific errors or deviations in detuning, amplitude and frequencies. Consequently, our scheme is analyzed for experimentally feasible configuration from source to detection, which introduces a different scheme for coherent amplification of SPP waves and should act as plasmonic field-effect transistor and surface plasmon lasers. 

\acknowledgments
SAZ acknowledges Barry C. Sanders and Klaus M\o lmer for their constructive discussions.
\bibliography{ref}
\begin{widetext}
\appendix
\section{\label{Section:Dynamics_Superradiant}Mathematical details of directional superradiant SPP formation}
In this section, we give a detailed explanation of exciting and launching directional SSPP emission, which is proportional to the excited $\ket{3}$ state within the interaction interface when the signal and driving field is switched off~(i.e. $\Omega_\text{s}=0$ and $\Omega_\text{d}=0$). Our SPP mode with a wavenumber $\bm{k}_\text{spp}$ propagates as superradiant emission through a collective excitation process if this single surface-plasmon field uniformly absorbed by an ensemble of $N_\text{a}$ atomic medium through time-Dicke state
\begin{equation}\label{timed-dicke_state}
    \ket{\psi_\text{SPP}}
    =\frac{1}{\sqrt{N_\text{a}}}
    \sum_{l=1}^{N_\text{a}}\exp\left[\text{i}
    \bm{k}_\text{SPP}\cdot \bm{r}_l\right]
    \ket{3_{i}}\otimes_{j\neq l}\ket{1_{j}}.
\end{equation}
The spontaneous emission from this prepared atomic ensemble, create a SPP field with wavevector $\hbar\bm{k}$, and the energy $\hbar(\omega_3-\omega_1)$. Similar to Ref.~\cite{PhysRevLett.96.010501}, this emission is superradiant and directional if $\bm{k}=\bm{k}_\text{SPP}$ and $\omega_\text{SPP}\approx\omega_{31}$. To satisfy these requirements, we suggest an optical pump and two contra-propagated couple laser fields as represented in Fig.~1 of the main text.

In our scheme, the optical pump is employed to induce coherence and hence provide collective excitation through $\ket{2}\leftrightarrow\ket{1}$ atomic transition and we employ $(2n_\text{p}+1)$-$\pi$ pulse to establish directionality. These pulse trains are resonant with $\ket{2}\leftrightarrow\ket{3}$ transition and their wavenumber $\bm{k}_{\text{C}\imath}$; $\imath\in\{1,2\}$ provides a unidirectional superradiant SPP wavenumber with~\cite{zhang2019surface}
\begin{equation}
  \bm{k}_\text{SPP}=(n_\text{p}+1)\bm{k}_{\text{C}2}-n_\text{p}\bm{k}_{\text{C}1}.  
  \label{Eq:Phase_Match_First_Appendix}
\end{equation}
As this SPP field coupled with the atomic state $\ket{3}$, the spectral component of the temporal atomic evolution then describes the superradiant surface-plasmonic dynamics.

Here we provide necessary steps toward SSPP dynamics. This evolution is the Fourier spectrum of the $c_{3}$ atomic state in the case that both signal and driving fields are switched off. We obtain the temporal evolution by exploiting the Schr\"odinger equation approach.

Finally, using this quantized electric field, considering the atomic dipole moment of the $\ket{3}\leftrightarrow\ket{1}$ as $\bm{d}_{l}$ and employing the Pauli matrices 
\begin{align}
    \sigma_{l}^{z}=&\ket{3_{l}}\bra{3_{l}}-\ket{1_{l}}\bra{1_{l}},\nonumber\\
    \sigma_{l}^{x}=&\ket{3_{l}}\bra{1_{l}}+\ket{1_{l}}\bra{3_{l}},
\end{align}
we achieve the Hamiltonian of the system as
\begin{equation}
    H=\sum_{l=1}^{N_\text{a}}\left[\frac{\hbar\omega}{2}\sigma_{l}^{z}-\sigma_{l}^{x}\bm{d}_{l}\cdot\bm{E}_{l}(\bm{r}_{l})\right]+\sum_{\jmath}\int\text{d}^3\bm{r}\int_{0}^{\infty}\text{d}\tilde{\omega}\hbar\tilde{\omega}\hat{\bm{C}}_{\jmath}^{\dagger}(\bm{r}',\omega)\cdot\hat{\bm{C}}_{\jmath}(\bm{r}',\omega).
    \label{Eq:Total_Hamiltonian_Appendix}
\end{equation}

Substituting Eqs.~\eqref{Eq:Total_Hamiltonian_Appendix} and \eqref{Eq:Initial_Ansatz} into Eq.~\eqref{Eq:Schrodinger_Equation}, mapping
\begin{equation}
    \frac{\mu_0\omega^2}{\pi}\bm{d}_\text{a}\cdot\mathcal{A}\cdot\bm{d}_\text{b}\sim g_\text{ab},\;\;\; \sum_{\jmath m}\int\text{d}^{3}\bm{r}\int_{0}^{\infty}\text{d}\tilde{\omega}\mapsto\int_\text{SPP},
\end{equation}
the temporal evolution of the atomic transitions are
\begin{align}
    \frac{\partial c_{1}(t)}{\partial t}=&\text{i}\sum_{l=1}^{N_\text{a}}c_{3}(t)g_\text{ab}^{*}(\bm{r}_\text{a},\bm{r}';\tilde{\omega})\text{e}^{\text{i}(\tilde{\omega}-\omega_{31})t},\label{Eq:Lower_state_Appendix}\\
    \frac{\partial c_{3}(t)}{\partial t}=&\text{i}\int_\text{SPP}c_{1}(\bm{r}',\tilde{\omega})g_\text{ab}(\bm{r}_\text{a},\bm{r}';\tilde{\omega})\text{e}^{-\text{i}(\tilde{\omega}-\omega_{31})t},\label{Eq:Excited_state_Appendix}
\end{align}
Next, we perform direct integration of the \eqref{Eq:Lower_state_Appendix} and plug the resultant equation into \eqref{Eq:Excited_state_Appendix}. Next, we employ
\begin{equation}
    g_\text{ab}(\bm{r}_\text{a},\bm{r}';\tilde{\omega})g_\text{ab}^{*}(\bm{r}_\text{a},\bm{r}';\tilde{\omega})=\frac{\mu_{0}^2\tilde{\omega}^4}{\pi}(\bm{d}_\text{a}\cdot\bm{\mathcal{A}}\cdot\bm{d}_\text{b})\times\left[\hbar\varepsilon_0\text{Im}[\varepsilon(\bm{r}',\tilde{\omega})]+\frac{\hbar^2\kappa_0^2}{\pi\tilde{\omega}^2}\bm{\nabla}\times\text{Im}[\kappa(\bm{r}',\omega)]\right],
\end{equation}
for $g_{31}:=g_\text{ab}g_\text{ab}^{*}$ and $\bar{\delta}=\tilde{\omega}-\omega_{31}$ to obtain the dynamics of the excited atomic states
\begin{equation}
    \frac{\partial c_3(t)}{\partial t}=-\sum_{l}\int_{0}^{\infty}\text{d}\tilde{\omega}\text{Im}[g_{31}]\int_0^{t}\text{d}\tau c_{3l}(\tau)\text{e}^{\text{i}\bar{\delta}(t-\tau)}.
    \label{Eq:First_Dynamics_Excited_State}
\end{equation}
Consequently, performing the integration over the frequency deviation $\tilde{\omega}$ and substituting into Eq.~\eqref{Eq:First_Dynamics_Excited_State} then yields an integro-differential equation that describes the dynamics of the excited atomic states.

We evaluate the emitter-emitter strength coupling~$g_{31}$ in a wavenumber space. To this aim, we assume the atoms are doped to the interface in a height $z_\text{at}$. Consequently, the SPP wave can couple to the atomic state only for $z\leq z_\text{at}$. The generated spontaneous emission, then produces a SPP field with an arbitrary wavenumber $\bm{q}$ and a coupling function
\begin{equation}
    g_{31}=\int\frac{\text{d}^2\bm{q}}{(2\pi)^2}g_\text{at}(\tilde{\omega};\bm{q})\text{e}^{\text{i}\bm{q}\cdot(\bm{r}_{l}-\bm{r}_{j})},
\end{equation}
similarly, we can represent the atomic-state amplitude in a wavenumber representation using a coupling function $\zeta^{(\text{C})}{(\text{C}}(\bm{q},\bm{k}_\text{SPP})=\braket{\Psi_{\bm{q}}|\Psi_{\bm{k}_\text{SPP}}}$.

Considering the dissipation of the SPP mode within the NIMM- quantum emitter interface as $\xi(\bm{q}):=\braket{\bm{q}|\bm{L}}:=\text{e}^{\text{i}\bm{q}\cdot\bm{L}}$, the dynamics of the excited atomic state becomes
\begin{equation}
    \frac{\partial c_3(t)}{\partial t}=-N_\text{a}\int\frac{\text{d}^2\bm{q}}{(2\pi)^2}\int\text{d}\tilde{\omega}\text{Im}[g_\text{at}]\zeta^{(\text{C})}{(\text{C}}(\bm{q},\bm{k}_\text{SPP})\xi(\bm{q})\times\int_0^{t}\text{d}\tau c_{3l}(\tau)\exp\{\text{i}(\tilde{\omega}-\omega_{31})(t-\tau)\}.
    \label{Eq:Excited_State_simplify}
\end{equation}
We assume the emitters are distributed within the interaction interface according to Gaussian distribution function
\begin{equation}
    \zeta^{(\text{C})}(\bm{q},\bm{k}_\text{SPP})=\exp\left\{\frac{-L^{2}(\bm{q}-\bm{k}_\text{SPP})^2}{2}\right\}.
    \label{Eq:Coupling_Function_Guass}
\end{equation}
We also evaluate the green tensor related to the interaction interface using the Fourier dynamics characterized by complex frequency and real wavenumber for a perturbation frequency $\tilde{\omega}=\omega_\text{SPP}-\text{i}\gamma_\text{SPP}$; $\gamma_\text{SPP}$ the total relaxation of the system, by expanding the approach represented in~\cite{PhysRevB.79.195414}.

Consequently, we define the reduced green tensor $\mathcal{A}_\text{SPP}(\bm{k}_\text{SPP})$ as the residue of the $g_\text{at}(\tilde{\omega},\bm{q})$ corresponds to pole $\tilde{\omega}=\omega_\text{SPP}-\text{i}\gamma_\text{SPP}$ and use this tensor to describe the atomic evolution. This green tensor for our hybrid interface depends on the optical properties of the metamaterial layer~$(\varepsilon_\text{N},\mu_{N})$, on propagation constant of dissipative $k_\text{N}$ and atomic medium $k_\text{a}$~($k_{\imath'}^{2}=k^2-\tilde{\omega}^2\varepsilon_{\jmath}\mu_{\jmath}/\text{c}^{2}$), on unit vector of the SPP field $u_\text{SPP}$ and is~\cite{PhysRevB.79.195414}
\begin{equation}
  \mathcal{A}_\text{SPP}=\text{i}\frac{k_\text{a}\varepsilon_\text{N}}{k_0\sqrt{\varepsilon_\text{a}}}\frac{C(k,\tilde{\omega})}{2\text{Re}[\omega_\text{SPP}]}\times\left[\bm{e}_{k}-\frac{k}{k_\text{N}}\bm{e}_{z}\right]\hat{\bm{u}}_\text{SPP}(\bm{r}_{l})\text{e}^{\text{i}(k_\text{N}-k_\text{a})z}. 
  \label{Eq:Green_Funtion_Interface_Appendix}
\end{equation}
In our calculation we exploit the parallel component to achieve the emitter-emitter coupling and ignore the dispersive transverse component. Assuming the Lorentzian lineshape for spectral distribution, we evaluate $\text{Im}[g_\text{at}]$ in \eqref{Eq:Excited_State_simplify} as
\begin{equation}
    \text{Im}[g_\text{at}]=\frac{\mathcal{A}_\text{SPP}^{\parallel}\gamma_\text{SPP}}{(\tilde{\omega}-\omega_\text{SPP})^2+\gamma_\text{SPP}^2}.
    \label{Eq:Green_Function_Residue}
\end{equation}
Finally, we substitute Eqs.~\eqref{Eq:Coupling_Function_Guass}, \eqref{Eq:Green_Function_Residue} and exploit $\text{d}^{2}\bm{q}=q\text{d}q\text{d}\phi$ to perform integration over $q$. Then we achieve
\begin{equation}
    -\frac{\partial c_{3}(t)}{\partial t}
    =\omega_\text{o}^2\exp\{(t_\text{L}\gamma_\text{SPP})^2\}\int_0^{t}\text{d}t'\;K(t,t')c_{3}(t'),
    \label{Eq:Volterra_Appendix}
\end{equation}
with $t_L=L/v_\text{SPP}$ is the SPP-flight time, $\tau_\text{l}=\gamma_\text{SPP}^{-1}$ is the time related to loss and $\omega_\text{o}^2:=(N_\text{a}/(2L)^2)g_\text{at}$ and with
\begin{equation}
    K(t,t')=\exp\left\{\frac{-(t-t'+2t_L^2\gamma_\text{SPP})^2}{(2t_L)^2}\right\}.
    \label{Eq:Kernel_Volterra}
\end{equation}
which is Eq.~\eqref{Eq:Volterra} in the main text. Note that in obtaining Eq.~\eqref{Eq:Volterra_Appendix}, we employ two valid and widely used assumptions, namely, (i) we treat the Schr\"odinger approach to describe the dynamics of the atomic ensemble, and (ii) we exploit the macroscopic Drude-Lorentz model to describe the optical frequency of our NIMM layer.

Using macroscopic model, we describe the electric permittivity~($\varepsilon_\text{N}$) and magnetic permeability~($\mu_\text{N}$) of the NIMM layer as
\begin{align}
    \varepsilon_\text{N}=&\varepsilon_{\infty}-\frac{\omega_\text{e}^2}{\omega_l(\omega_l+\text{i}\gamma_\text{e})},\\
    \mu_\text{N}=&\mu_{\infty}-\frac{\omega_\text{m}^2}{\omega_l(\omega_l+\text{i}\gamma_\text{m})},
\end{align}
for~$\varepsilon_{\infty}$ and~$\mu_{\infty}$ the background constant for the permittivity and permeability, respectively. The other constants are $\omega_l$ the perturbation frequency, $\omega_\text{e}$~($\omega_\text{m}$) are the electric and magnetic plasma frequencies, and $\gamma_\text{e}$~($\gamma_\text{m}$) are the corresponding decay rates. Specifically, we assume a nano-fishnet metamaterial layer fabricated with $\text{Al}_2\text{O}_3$-$\text{Ag}$-$\text{Al}_2\text{O}_3$ multilayer with rectangular nano-hole structure, which according to Ref.~\cite{Xiao:09} provides SPWs within optical frequency region. We exploit these parameters to describe the NIMM layer: $\varepsilon_{\infty}=\mu_{\infty}=1.2$, $\omega_\text{e}=1.37\times10^{16}~\text{s}^{-1}$, $\omega_\text{m}=10^{15}~\text{s}^{-1}$, $\gamma_\text{e}=2.37\times10^{13}~\text{s}^{-1}$ and $\gamma_\text{m}=10^{12}~\text{s}^{-1}$. This waveguide is low-loss for the for our $\ket{3}\leftrightarrow\ket{1}$ (see main text for details of the realistic atomic medium and corresponds parameters) transition wavelength. Plugging into Eqs.~\eqref{Eq:Green_Funtion_Interface_Appendix} and~\eqref{Eq:Green_Function_Residue}, we can obtain the dynamics of the superradiant SPP dynamics within quantum emitter-NIMM layer interface when the signal and driving field is switched off.

\section{\label{Sec:Surface_plasmon_amplification_Appendix} Surface-plasmon amplification by superradiant emission of radiation}
In this section, we provide mathematical details of the surface plasmon amplification in the presence of the directional superradiant surface-plasmon polariton launching. We proceed this section and elucidate the details of derivation in two subsections: In subsection \S~\ref{Dynamics_weak_SPP} we discuss the mathematical details of the weak signal field dynamics and in subsection \S~\ref{Maxwell_Schrodinger_Amplification_SPP} we give the temporal evolution of the atomic medium in the presence of the signal and driving laser fields. Our approach for obtaining the spatiotemporal dynamics of weak surface-plasmon polariton field is achieved by employing the Fourier analysis commensurate with the Heisenberg equation of motion and we investigate the evolution of the atomic medium by using the Maxwell-Schr\"odinger equations, which are valid assumptions.

\subsection{\label{Dynamics_weak_SPP} Mathematical details of signal field dynamics}
In this subsection, we provide the mathematical details of the spatiotemporal evolution of the weak probe pulse and its amplification within atomic medium-NIMM layer interface. As it is shown in the main text, our amplification scheme is achieved by coupling a strong driving field, and two contra-propagating externally added driving fields, which are employed to suppress unwanted Stark shifts and reduced Ohmic loss, respectively. The weak SPP pulse is hence ultra low-loss and propagate through the interaction interface.

We achieve the dynamics of the weak SPP field, defined as $\Omega_\text{s}:=\bm{d}_\text{l}\cdot\hat{\bm{E}}(\bm{r},t)$ in a wavenumber representation as
\begin{equation}
    \tilde{\Omega}_\text{s}\approx\sum_{\jmath,m}\int\frac{\text{d}^2\bm{q}}{(2\pi)^2}\int\text{d}\tilde{\omega}\tilde{\mathcal{A}}_{13,m}C_{\jmath m}(\bm{q},\omega)\text{e}^{\text{i}(\bm{q}\cdot\bm{r}-\tilde{\omega}t)}.
    \label{Eq:Fourier_Dynamics}
\end{equation}
To perform the integration, first we should evaluate the dynamics of the quantized signal field $\hat{\bm{C}}_{\jmath}^{\text{s}}$ in a wavenumber representation using the Heisenberg equation of motion. Also, this integration vanishes for the large perturbations in wave-vector and frequencies. Therefore for $\bm{q}\gg\bm{\mathcal{K}}$~($\bm{q}\ll\bm{\mathcal{K}}$); $\mathcal{K}$ the dispersion of the signal plasmonic field, the signal SPP is highly dissipative due to atomic absorption. Consequently, we consider the resonant coupling between this plasmonic field and the atomic medium. To this aim, we choose
\begin{align}
    \bm{q}=&\bm{k}_\text{SPP}+\mathcal{O}(\bm{q-\bm{k}_\text{SPP}}),\label{Eq:Perturbed_Wavenumber}\\ 
    \omega=&\omega_\text{SPP}+v_\text{SPP}\delta\omega+\mathcal{O}(\delta\omega^{2}),
    \label{Eq:Perturbed_Frequency}
\end{align}
and assume $\omega:=\omega_\text{SPP}+\text{i}\gamma_\text{SPP}$, $\mathcal{K}\approx k_\text{SPP}$.

Now we can evaluate the plasmonic signal field in a Fourier space representation. Plugging Eqs.~\eqref{Eq:Perturbed_Wavenumber} and~\eqref{Eq:Perturbed_Frequency} into Eq.~\eqref{Eq:Fourier_Dynamics}, performing integration over $\tilde{\omega}$ by using $\delta\tilde{\omega}\ll\omega_\text{SPP}$, defining $\tilde{\mathcal{A}}_{13,m}^{(\text{s})}=\mathcal{A}_{13,m}(\omega_\text{SPP},\mathcal{K})$ and $\delta\bm{q}=\bm{q}-\bm{\mathcal{K}}$, we achieve
\begin{equation}
    \tilde{\Omega}_\text{s}\approx\sum_{\jmath,m}\int\frac{\text{d}^2\bm{q}}{(2\pi)^2}\tilde{\mathcal{A}}_{13,m}^{(\text{s})}C_{\jmath m}(\bm{q},\omega)\text{e}^{\text{i}(\delta\bm{q}\cdot\bm{r}-v_\text{SPP}t)}.
    \label{Eq:Simplified_Fourier_Signal_SPP}
\end{equation}
To achieve the spatiotemporal dynamics of \eqref{Eq:Simplified_Fourier_Signal_SPP}, first we investigate the dynamics of the $\hat{\bm{C}}_{\jmath}$ within quantum emitter-metamaterial interface. Considering the Hamiltonian of the system \eqref{Eq:Total_Hamiltonian}, we achieve the equation of motion for $\hat{\bm{C}}_\jmath$ in terms of the atomic ground state $\ket{G}$ and excited state $\ket{\Psi_0}$ using the Heisenberg approach and defining~\footnote{The excited atom in this case is placed at $\bm{r}=\bm{r}_{l}$}
\begin{equation}
    s(t):=\bra{G}\sigma_{l}^{-}\ket{\Psi_0}+\bra{\Psi_0}\sigma_{l}^{+}\ket{G}
\end{equation}
as
\begin{equation}
    \text{i}\frac{\text{d}C_{\jmath,m}}{\text{d}t}=\tilde{\omega}C_{\jmath,m}-\sum_{l=1}^{N_\text{a}}g_\text{at}^{*}(\bm{r_{l}},\bm{r}';\tilde{\omega})\left[s(t)+s^{*}(t)\right],
    \label{Eq:Dynamics_Annihilation}
\end{equation}
with $s(t)+s^{*}(t)\sim \rho_{31}(t)$. Finally, we perform the derivative with respect to $\bm{r}'$ and time $t$ from Eq.~\eqref{Eq:Simplified_Fourier_Signal_SPP} and make use of Eq.~\eqref{Eq:Dynamics_Annihilation} to achieve
\begin{equation}
    \left(\frac{\partial}{\partial t}+\bm{v}_\text{SPP}\cdot\bm{\nabla}\right)\Omega_\text{s}=\text{i}\mathcal{C}\tilde{\rho}_{31},
   \label{Eq:Main_equation_Appendix}
\end{equation}
which is a Maxwell-Schr\"odinger equation for our plasmonic system. Eq.~\eqref{Eq:Main_equation_Appendix} describe the evolution of the weak SPWs within a hybrid atomic medium-metamaterial interface.

\subsection{\label{Maxwell_Schrodinger_Amplification_SPP}Maxwell-Schr\"odinger equation and amplification of weak SPP field}
In this subsection, we present the steps that yield the amplification of weak signal polaritonic field. To this aim, we solve the Schr\"odinger equation to achieve the dynamical evolution of the atomic dipole moment $\tilde{\rho}_{31}$ and we exploit this quantity to describe the amplification of SPP field using Eq.~\eqref{Eq:Main_equation_Appendix}. Our approach for coherent amplification is based on Maxwell-Schr\"odinger equations, which is a well-established and valid assumption.

In the presence of strong driving and weak signal field the total electric field of the system is
\begin{equation}
    \bm{E}_\text{T}=E_\text{d}\bm{u}_\text{d}(z)+E_\text{s}\bm{u}_\text{s}(z).
    \label{Eq:Total_Electric_Field}
\end{equation}
To efficient amplification of the weak SPP field, one suggestion is to couple the signal field to $\ket{3}\leftrightarrow\ket{1}$ transition, and couple driving field to both $\ket{3}\leftrightarrow\ket{1}$ and $\ket{2}\leftrightarrow\ket{1}$ transitions using an intermediate state $\ket{a}$~\cite{PhysRevX.3.041001}. We represent the unit vector of the $\ket{i}\leftrightarrow\ket{j}$ atomic transition as $\bm{e}_{ij}$~($\bm{d}_{ij}=|\bm{d}_{ij}|\bm{e}_{ij}$), define the confinement of the field to the interface as
\begin{align}
    \zeta_\text{s}(z)=&\bm{u}_\text{s}(z)\cdot\bm{e}_{31},\\
    \zeta_\text{d}^{(1)}(z)=&\bm{u}_\text{d}(z)\cdot\bm{e}_{31},\;\;\;\zeta_\text{d}^{(2)}(z)=\bm{u}_\text{d}(z)\cdot\bm{e}_{a1},
\end{align}
and the Rabi frequencies for these fields as
\begin{align}
    \Omega_\text{s}=&\frac{E_\text{s}d_{31}}{\hbar},\\
    \Omega_\text{d}^{(1)}=&\frac{E_\text{s}d_{31}}{\hbar},\\ \Omega_\text{d}^{(2)}=&\frac{E_\text{s}d_{a1}}{\hbar}.
\end{align}
The Hamiltonian of the total system is
\begin{align}
    H_\text{I}=\hbar\left[\zeta_\text{d}^{(1)}(z)\Omega_\text{d}^{(1)}\text{e}^{-\text{i}\omega_{31}t}\ket{3}\bra{1}+\zeta_\text{d}^{(2)}(z)\Omega_\text{d}^{(2)}\text{e}^{-\text{i}\omega_{a1}t}\ket{a}\bra{1}+\zeta_\text{d}^{(1)}(z)\Omega_\text{s}\text{e}^{-\text{i}\omega_{31}t}\ket{3}\bra{1}+\text{c}.\text{c}.\right],
\end{align}
and we assume the relaxation rate of the $\ket{1}$~($\ket{3}$) is $\gamma_{1}$~($\gamma_{3}$), respectively. We Consider the ansatz as
\begin{equation}
    \ket{\psi(t)}=c_{1}(t)\ket{1}+c_{a}(t)\ket{a}+c_{3}(t)\ket{3}.
\end{equation}
The atomic medium and correspond atomic transitions are affected by the deriving and signal plasmonic fields. We employ mapping for coupling fields~\cite{PhysRevA.98.013825} to express the atomic states as
\begin{align}
    \braket{\zeta(z)\Omega_l}\mapsto&\Omega_l,\\
    c_\text{3}\text{e}^{-\text{i}\omega_{31}t}\mapsto&\tilde{c}_{3},\; c_\text{1}\text{e}^{-\text{i}\omega_{\text{a},1}t}\mapsto\tilde{c}_{1},
\end{align}
and consequently achieve the temporal evolution of the atomic states
\begin{align}
    \dot{c}_{1}(t)+\text{i}\Omega_\text{d}^{(1)}c_{3}(t)+\text{i}\Omega_\text{d}^{(2)}c_{a}(t)-\frac{\gamma_\text{s}}{2}c_{3}-\frac{\gamma_\text{a}}{2}c_{a}=&0,\label{Eq:Ground_State_Total_Appendix}\\
    \dot{c}_{3}(t)+\left(\text{i}\omega_{31}+\frac{\gamma_3}{2}\right)c_{3}(t)+\text{i}\Omega_\text{d}^{(1)}c_{3,\text{s}}+\text{i}\Omega_\text{d}^{(2)}c_{1,\text{d}}=&0.\label{Eq:Excited_State_Total_Appendix}
\end{align}

These atomic states are coupled to a signal field component with frequency $\omega_{31}\approx\omega_\text{SPP}$ and thereby the temporal distribution of these states are
\begin{equation}
    c_{i}=c_{i,\text{s}}\text{e}^{\text{i}(\omega_\text{SPP}t-\mathcal{K}\cdot\bm{r})-\gamma_\text{SPP}t}+\text{c}.\text{c}.,\;\; i\in\{1,3\}.
    \label{Eq:Excited_State_Modulation}
\end{equation}
Plugging Eq.~\eqref{Eq:Excited_State_Modulation} into Eqs.~\eqref{Eq:Ground_State_Total_Appendix} and \eqref{Eq:Excited_State_Total_Appendix}, we achieve the dynamical evolution of the atomic states for both driving field and weak signal field modulations. In our analysis, we neglect the temporal variation of the ground state due to coupling with a weak plasmonic signal field $\dot{c}_{1,\text{s}}\ll1$. Consequently, the density matrix elements in \eqref{Eq:Main_equation_Appendix} is
\begin{equation}
    \tilde{\rho}_{31}=\left[c_{3,\text{s}}^{*}(t)c_{1,\text{d}}(t)+c_{3,\text{d}}^{*}(t)c_{1,\text{s}}(t)\right]\text{e}^{-\text{i}\theta-\gamma_\text{SPP}t},
    \label{Eq:Atomic_Dipole_Final_Appendix}
\end{equation}
here $\varphi:=\omega_\text{SPP}t-\bm{\mathcal{K}}\cdot\bm{r}$. In order to obtain the dynamics of the amplified SPP field, we plug Eq.~\eqref{Eq:Atomic_Dipole_Final_Appendix} into~\eqref{Eq:Main_equation_Appendix}, perform a time derivative with respect to $t$ and replace the temporal atomic evolution from Eqs.~\eqref{Eq:Ground_State_Total_Appendix} and \eqref{Eq:Excited_State_Total_Appendix}, we achieve
\begin{equation}
   \left(\frac{\partial}{\partial t}+\beta(x,t)\right)\left(\frac{\partial}{\partial t}+\bm{v}_\text{SPP}\cdot\bm{\nabla}\right)\Omega_\text{s}=\text{i}\xi f(x,t)\Omega_\text{s},
   \label{Eq:Main_equaiton_Amplification_Appendix}
\end{equation}
with the coefficients $\beta$ and $f$ defined as
\begin{align}
    \beta:=&\tilde{\Delta}+\frac{\bar{\Omega}_\text{d}^{(1)2}+\bar{\Omega}_\text{d}^{(2)2}}{-\text{i}\omega_\text{SPP}+\gamma_\text{SPP}}+\frac{4\bar{\Omega}_\text{d}^{(1)}\bar{\Omega}_\text{d}^{(2)*}\cos(2\varphi)}{\omega_\text{SPP}+\text{i}\gamma_\text{SPP}},\nonumber\\
    f:=&1-\frac{2\bar{\Omega}_\text{d}^{(1)2}}{\tilde{\Delta}(-\text{i}\omega_\text{SPP}+\gamma_\text{SPP})}-\frac{2\bar{\Omega}_\text{d}^{(1)2}+\bar{\Omega}_\text{d}^{(2)2}\exp\{2\text{i}\varphi\}}{\omega_\text{SPP}(-\text{i}\omega_\text{SPP}+\gamma_\text{SPP})}.
\end{align}
We employ Eq.~\eqref{Eq:Main_equaiton_Amplification_Appendix} to describe the coherent amplification of the weak SPP field.

Note that our quantitative description of the system and deriving Eq.~\eqref{Eq:Main_equaiton_Amplification_Appendix} is based on three assumptions: (i) employing the Sch\"odinger equation to achieve atomic ensemble dynamics~\cite{scully1999quantum}, (ii) exploiting Drude-Lorentz model~\cite{PhysRevLett.101.263601,Xiao:09} to describe nano-fishnet metamaterial layer, and (iii) using Maxwell-Schr\"odinger equation~\cite{PhysRevX.3.041001} to achieve the dynamics of the weak surface-plasmon polariton field. These are well-established methods to describe a nano-optic configuration and justify the validity of our theoretical model.

\section{\label{Sec:Calculation_far_Field_Superradiant}Calculation of the far-field superradiant field within interaction interface}
In this section, we present the mathematical steps towards superradiant emission of radiation. We treat this superradiant plasmonic as far-field and employ the Dyadic green tensor to characterize this plasmonic field. We assume $X_{\mu\nu}$ to be this green tensor and characterize its components as
\begin{align}
    X_{xx}(\bm{k}_{\parallel},z')&:=\text{i}\frac{\mathcal{K}}{2k_{2}^{3}}\exp\{\text{i}\mathcal{K}|z_\text{at}-z|\},\\
    X_{xz}(\bm{k}_{\parallel},z')&:=\text{i}\frac{k_{\parallel}}{2k_{2}^{3}}\exp\{\text{i}\mathcal{K}|z_\text{at}-z|\},\\
    X_{yy}(\bm{k}_{\parallel},z')&:=\text{i}\frac{1}{2\mathcal{K}}\exp\{\text{i}\mathcal{K}|z_\text{at}-z|\},
\end{align}
$X_{xz}=X_{zx}$, $\tilde{g}_{\mu\nu}:=g_{z_\text{at},\mu\nu}$ and considering
\begin{equation}
    \bm{k}_\parallel=k_{x}\bm{e}_{x}+k_{y}\bm{e}_{y}
\end{equation}
as 
\begin{align}
    \tilde{g}_{xx}&=\frac{1}{|\bm{k}_{||}|^{2}}\exp\{-\text{i}\mathcal{K}(z+z_\text{at})\}\left[-k_{x}^{2}r_\text{p}^{k_{x}}X_{xx}(\bm{k}_{\parallel},z')+k_{y}^{2}r_\text{s}^{k_{y}}X_{yy}(\bm{k}_{\parallel},z')\right],\\
    \tilde{g}_{xy}&=-\frac{k_{x}k_{y}}{|\bm{k}_{||}|^{2}}\exp\{-\text{i}\mathcal{K}(z-z_\text{at})\}\left[r_\text{p}^{k_{x}}X_{xx}(\bm{k}_{\parallel},z')+r_\text{s}^{k_{y}}X_{yy}(\bm{k}_{\parallel},z')\right],\;\;\;\;\;\tilde{g}_{xy}=\tilde{g}_{yx},\\
    \tilde{g}_{xy}&=\frac{1}{|\bm{k}_{||}|^{2}}\exp\{-\text{i}\mathcal{K}(z+z_\text{at})\}\left[-k_{y}^{2}r_\text{p}^{k_{y}}X_{xx}(\bm{k}_{\parallel},z')+k_{x}^{2}r_\text{s}^{k_{x}}X_{yy}(\bm{k}_{\parallel},z')\right],\\
    \tilde{g}_{xz}&=-\frac{k_{x}}{|\bm{k}_{||}|}r_\text{p}^{k_{x}}X_{xz}(\bm{k}_{\parallel},z')\exp\{-\text{i}\mathcal{K}(z-z_\text{at})\},\;\tilde{g}_{zx}=\frac{k_{x}}{|\bm{k}_{||}|}r_\text{p}^{k_{x}}X_{xz}(\bm{k}_{\parallel},z')\exp\{-\text{i}\mathcal{K}(z-z_\text{at})\},\\
    \tilde{g}_{yz}&=-\frac{k_{y}}{|\bm{k}_{\parallel}|}r_\text{p}^{k_{y}}X_{xz}(\bm{k}_{\parallel},z')\exp\{-\text{i}\mathcal{K}(z-z_\text{at})\},\;\tilde{g}_{zx}=\frac{k_{y}}{|\bm{k}_{||}|}r_\text{p}^{k_{y}}X_{xz}(\bm{k}_{\parallel},z')\exp\{-\text{i}\mathcal{K}(z-z_\text{at})\},\\
    \tilde{g}_{yz}&=\frac{|k_{\parallel}|}{\mathcal{K}}r_\text{p}^{k_{x}}X_{xz}(\bm{k}_{\parallel},z')\exp\{-\text{i}\mathcal{K}(z-z_\text{at})\}.
\end{align}
Next, we define the polar coordinate in terms of azimuth~($\phi$) and polar $\theta$ angles
\begin{equation}
    \bm{e}_{r}:=\sin\theta\left(\cos\phi\bm{e}_{x}+\sin\phi\bm{e}_{y}\right),
\end{equation}
we obtain the green tensor for a far-field superradiant SPP as
\begin{equation}
    g_{\mu\nu}(\bm{r},\bm{r'},\mathcal{K};\omega)=\text{i}\frac{1}{2\pi}\left(\frac{\mathcal{K}z}{r^{2}}\right)\exp\left\{\text{i}\bm{\mathcal{K}}\cdot\left(\bm{r}-\bm{r}'\right)\right\}\times\tilde{g}_{\mu\nu}(\mathcal{K}\sin\theta\cos\phi,\mathcal{K}\sin\theta\sin\phi,\omega;z').
\end{equation}
The electric field due to this far-field dyadic tensor is achieved by performing the integration over all possible times. Taking into account the relaxation rates of the atomic ensemble, this electric field relates to the phasing term
\begin{equation}
    |\bm{S}|\sim|\tilde{g}_{\mu\nu}(\mathcal{K}\sin\theta\cos\phi,\mathcal{K}\sin\theta\sin\phi,\omega;z')|^{2}\times\left|\sum_{j}^{N_\text{a}}\exp\{\text{i}\mathcal{K}\sin\theta(\cos\phi x+\sin\phi y)-\gamma_\text{R}t\}\right|^{2},
\end{equation}
indicates that the intensity pattern of the far-field superradiant SPP exist for the $k_\text{SPP}$ satisfying the phase-match condition and also constructive interference of the superradiant field phase. These two conditions provides unique solution for the superradiant SPP intensity pattern that stands for \textit{unidirectional} superradiant SPP generation.

\end{widetext}

\end{document}